# COMMUNICATION SYSTEMS USING LABVIEW

Samir Katte UID: 304149393
Advisor: Prof. Gregory Pottie

# **Introduction**

The following report contains user guides of four modulation schemes implementation using LabVIEW. The list of various modulation schemes is as follows:

- Amplitude Modulation
- Frequency Modulation
- Binary Frequency Shift Keying
- Quadrature Amplitude Modulation

This report was submitted for the complete fulfillment of the EE299 MS project course required for the Master of Science in Electrical Engineering degree at the University of California-Los Angeles in the project option.

# **Acknowledgements**

I would like to express my sincere thanks to Professor Gregory Pottie for his valuable guidance and constant influx of ideas while implementing this project. A special thanks to Kyle Mozdzyn from National Instruments, for his suggestions and technical assistance during the course of this project.

# Amplitude Modulation using LabVIEW

## User Guide

## Introduction

LabVIEW enables engineers to simulate various communication and control systems. LabVIEW helps to create Virtual Instruments (VIs) which are the files with which the user interacts to accomplish the required task.

The AM system is implemented using two separate VIs i.e. Transmitter_AM.vi and Receiver_AM.vi. Each VI has two parts: Front Panel and the Block Diagram. The Front Panel is usually the interface the user interacts with and observes results. The block diagram contains the blocks used to implement the functionality required for the operation of the VI. The individual blocks in the block diagram are called the sub VIs. The user may or may not need to make changes in the block diagram of the VI during the execution of the LabVIEW program.

## Hardware setup:

1. Make sure you have LabVIEW installed on your machine so that you can run the LabVIEW VIs on it.
2. Use a headphone as a transmitter for the AM wave.
3. The microphone of the machine will be used as the receiver. It is advisable to keep the headphone near the microphone for better results. It is also suggested to have a quiet environment during the whole experiment.

## AM Transmitter Working Principle

Consider a sinusoidal carrier wave defined by $c(t)=A_c \cos(2\pi f_c t)$ where $A_c$ is the carrier amplitude and $f_c$ is the carrier frequency.

Amplitude modulation (AM) is defined as a process in which the amplitude of the carrier wave c(t) is varied about a mean value, linearly with message signal m(t). An AM wave may be described as follows:

$$s(t) = A_c [1+ k_a m(t)] \cos(2\pi f_c t)$$

where $k_a$ is a constant called amplitude sensitivity (its value should lie between 0 and 1) of the modulator responsible for the generation of the modulated signal s(t). It should be ensured that $|k_a m(t)|<1$ and that $f_c >>W$ where W is the message bandwidth for effective modulation.

The following diagram shows how a high frequency carrier wave is modulated by a low frequency message signal

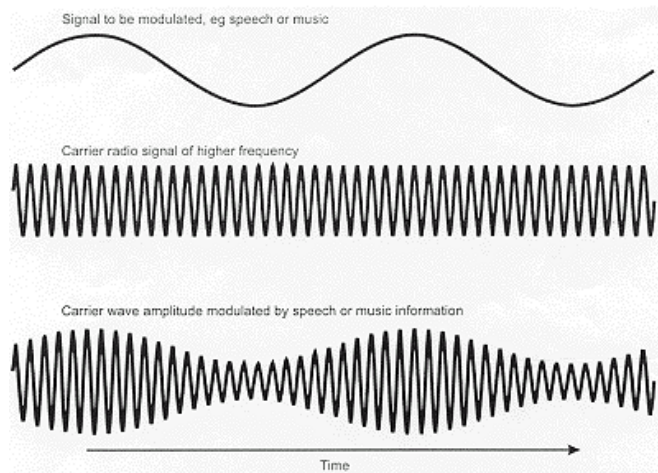

The frequency spectrum is seen as follows where the upper spectrum is of the message signal and the lower one is of the amplitude modulated wave.

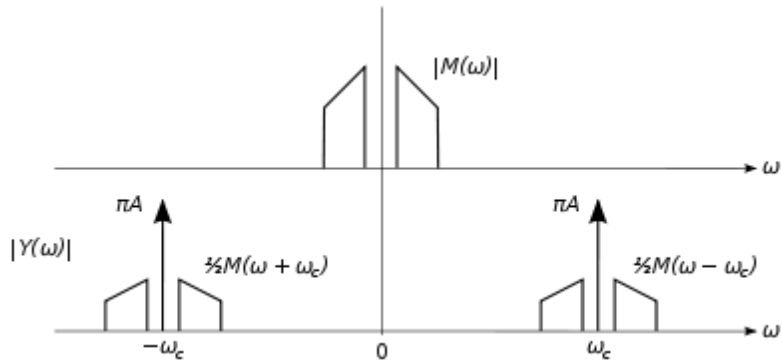

## Transmitter block

The transmitter block diagram looks as follows:

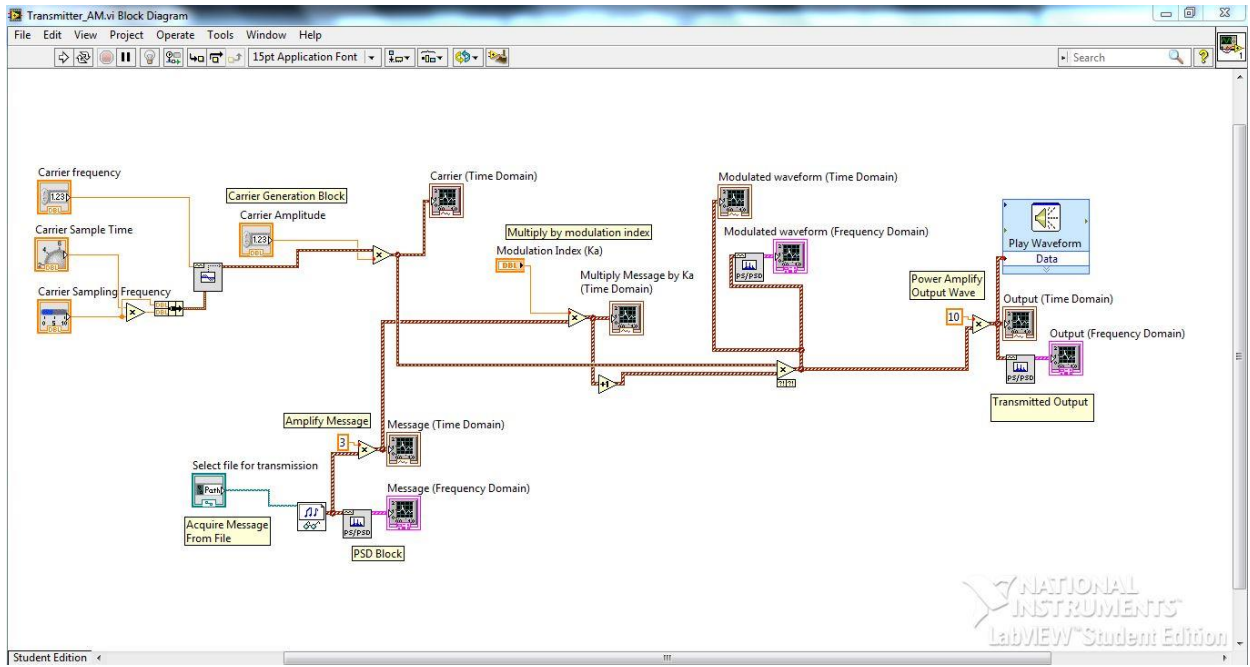

We all know that Amplitude Modulation consists of changing the amplitude of the high frequency carrier wave with a low frequency message signal. The above VI does the same by using a .wav file as a message.

The key functions in the block diagram are labeled in yellow. I will explain them one by one. Pay keen attention to the things to be edited in the front panel and the block diagram. Most of the blocks mentioned below are the only ones which can be used for that specific application in LabVIEW. Some suggestions for using different blocks are mentioned at particular places in the description below.

1. **Carrier Generation Block:** The carrier frequency, sample time and the carrier sampling frequency are the variables that can be adjusted by using the controls created on the front panel. Remember that the sampling frequency should be more than twice the carrier frequency according to Nyquist criterion. The sample time will decide the number of seconds for which the carrier is generated. You will usually use a message signal of 10 seconds, so set the sampling time to 10 seconds.

Another block which can be used for a sine wave generation is a Sine Wave sub VI block.

2. **Acquire Message from File:** In the space corresponding to this sub VI on the front panel, enter the path where the message file is stored. Make sure that the file is in the '.wav' format else LabVIEW will reject it.

   In the block diagram, a constant is used for multiplication which can be edited. It can be replaced by a control, so that it can be adjusted from the front panel. To replace it, delete the constant input to it and right click on the sub VI input pin, and select 'create control' option.

3. **Amplify Message:** The tiny block showing '3' can be edited by the user in the block diagram to amplify the message. Care should be taken to avoid the multiplication of the message and the modulation index ($k_a$) result in a waveform with values greater than 1 or lesser than -1. In short $|m*k_a|<1$, where m is the message signal. By observing the corresponding graph on the Front Panel, you can ensure that the condition is met and you can change the message amplification accordingly

4. **PSD block:** It converts the time signal into a frequency spectrum data. It is displayed on the graph. Similar blocks can be observed in the entire block diagram. You can also add your own time signal graph blocks or PSD blocks wherever in the block diagram to observe the waveform at that point of the system. Don't forget to position the corresponding graph on the front panel by dragging and adjusting its size.

5. **Multiply by Modulation Index:** Enter a value between 0 and 1 in the corresponding space in the front panel.

   Notice that the value is a number of type 'double' in this example. We can use other options to introduce it in the diagram as mentioned in the description above.

6. **Power Amplify Output Wave:** Presently the output wave is power amplified by a factor of 10. You can change it if you want to by clicking in the small box in the block diagram.

The usual sampling frequency should be 44100 or 22050 because these values of sampling frequencies are compatible with most of the audio files and sound card drivers. Using these sampling values will prevent us from using resample waveform VI blocks in our design. The resample waveform block takes a signal and a 'dt' value as the inputs and it outputs the resampled waveform. The value 'dt' is the reciprocal of the sampling frequency desired.

**The front panel looks as follows**

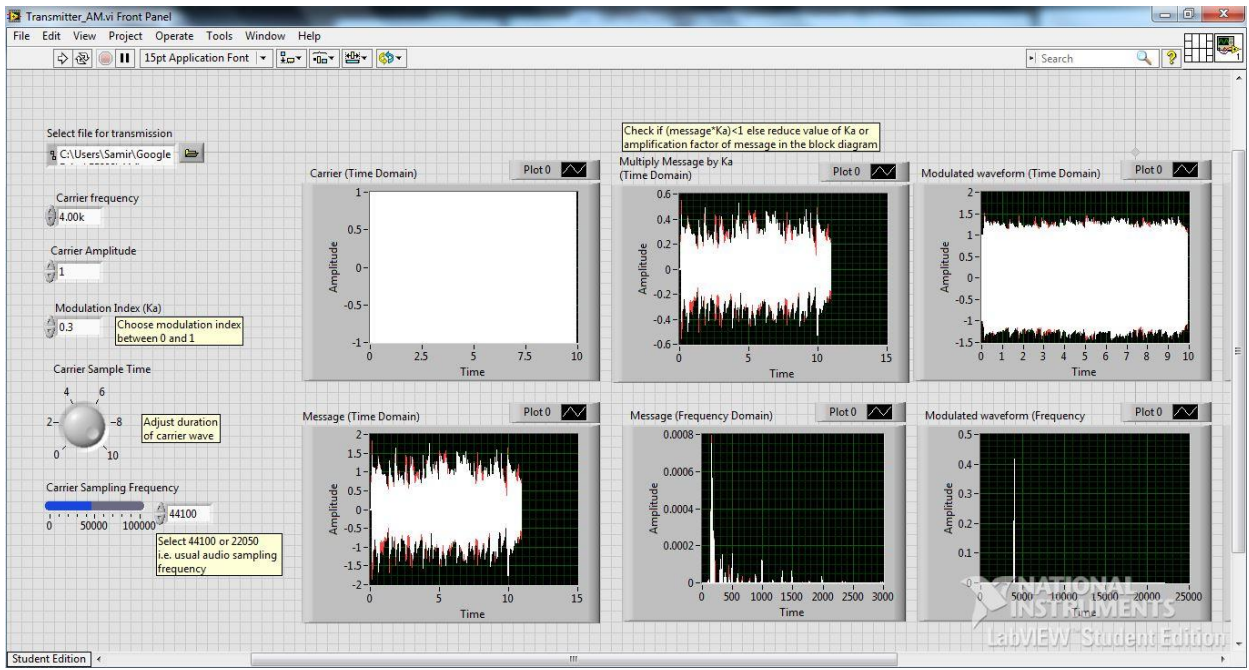

Most of the adjustments to be made in the front panel are explained above. My settings are 4kHz carrier with amplitude 1volt, modulation index=0.3, carrier sample time =10 seconds for a 10 second message wave and 44100 as the carrier sampling frequency, so that the carrier is compatible with the sound card driver.

There is one more important setting to be set. It is in the 'Play Waveform' sub VI. When we use the 'Play Waveform' for the first time, you need to select the output device i.e. speakers or headphones. Test device to hear the audible tone. If you don't hear it, select another device. Rest of the settings fairly remains the same.

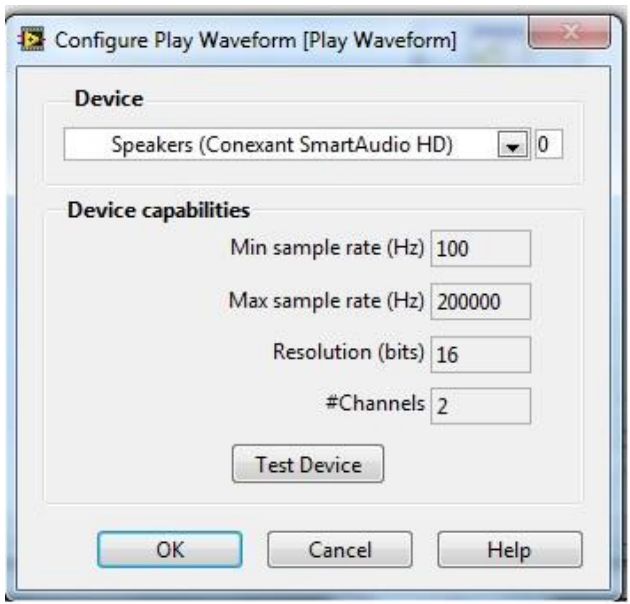

## AM Receiver Working Principle

The demodulation of an AM wave can be accomplished by means of a simple circuit called the envelope detector.

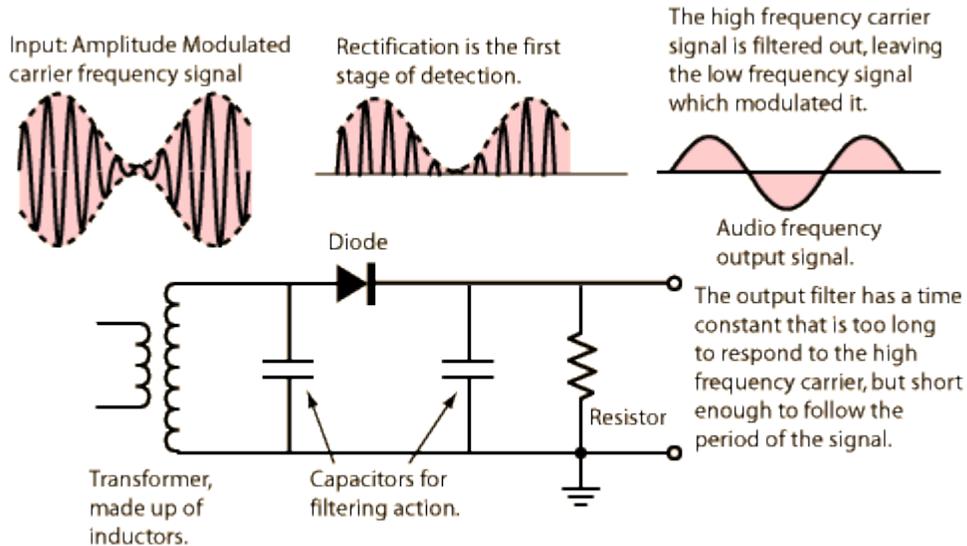

The envelope detector consists of a diode and resistor-capacitor (RC) filter. During the positive half cycle, the diode is forward biased and it charges the capacitor. Once the capacitor is charged at a voltage higher than the input to the diode, the diode gets reversed biased and the capacitor gets discharged through the resistor. In short, the diode acts as a rectifier and the RC filter removes the high frequency ripple from the signal. Hence the output is the low frequency message signal.

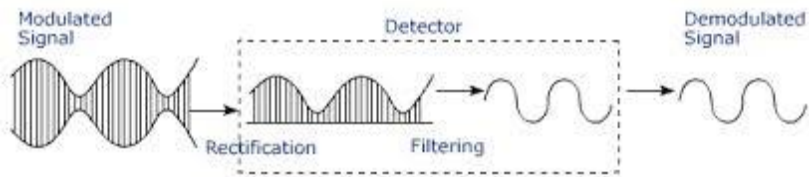

## Receiver Block

The receiver block looks as shown below

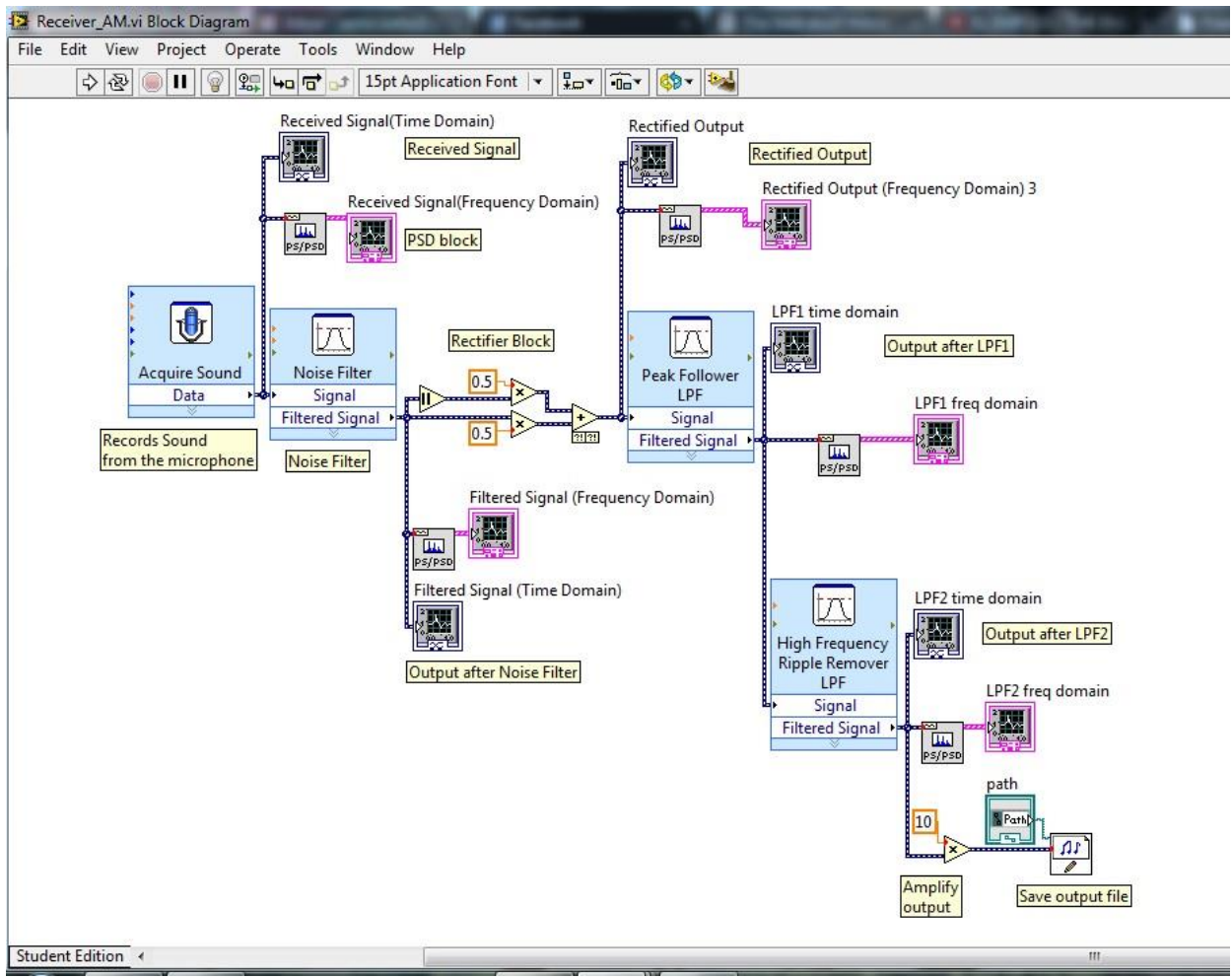

The important components labeled in yellow are explained below:

1. **Record Sound from the Microphone:** This sub VI is used to record the sound from the microphone.

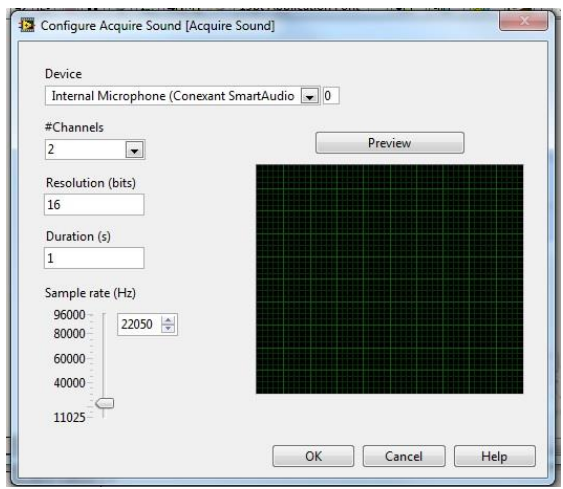

Select the input device i.e. microphone which you are using and also specify the number of seconds you want to record the sound for. For an input message signal of 10 seconds, you should keep the duration in the 'Acquire Sound' sub VI as 10 seconds too.

2. **Noise Filter:** Filters are used extensively in all communication system implementations in LabVIEW. This is how the filter sub VI looks like.

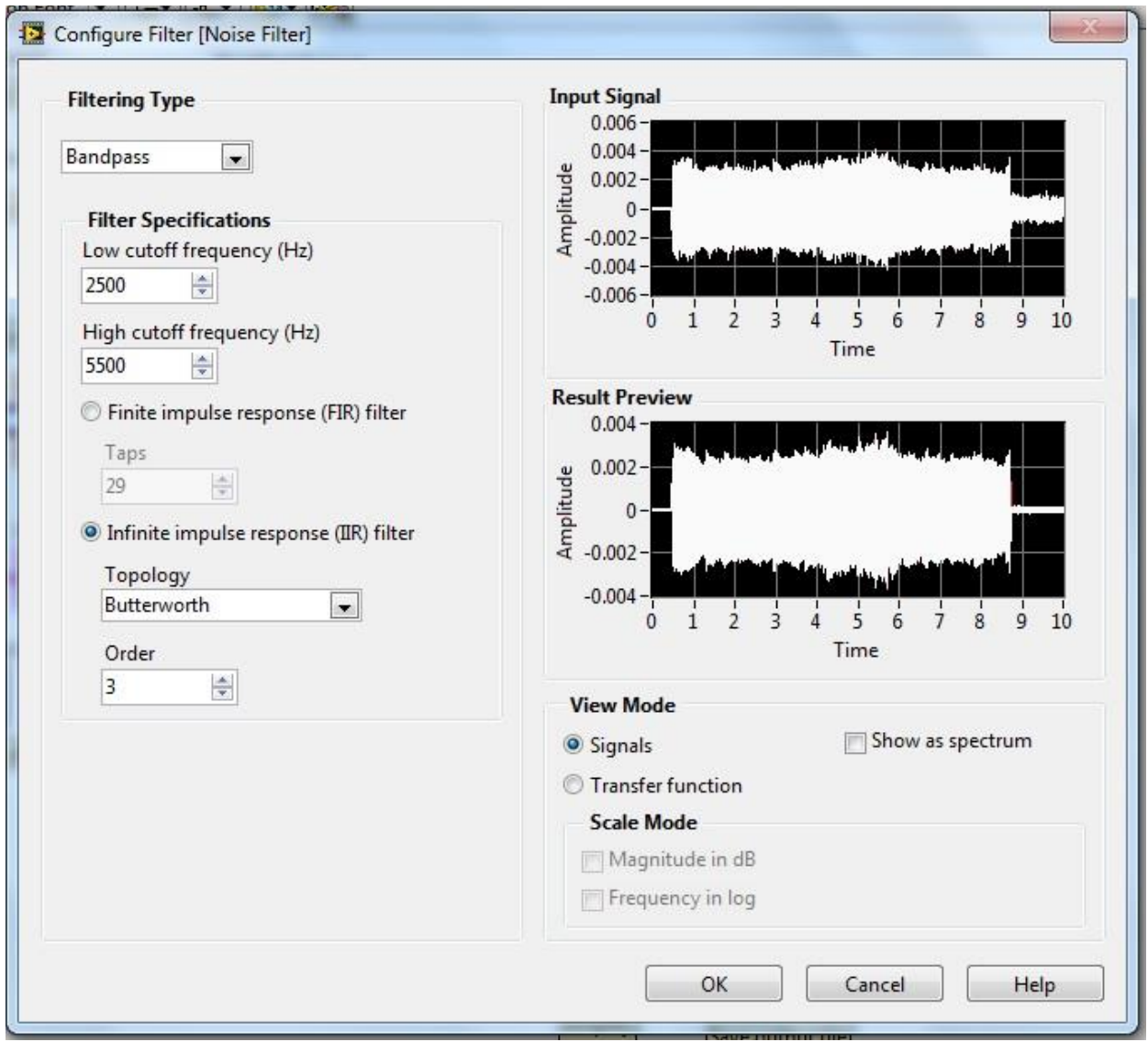

You can select the filter type and the cutoff frequency as well as the order of the filter. I used a bandpass filter with a bandwidth of 3kHz around the carrier frequency to eliminate noise and because the message sound file had a spectrum from 0Hz to 1500Hz.

3. **Rectifier block and Filters:** We know how the AM peak detector circuit looks like. The diode acts as a rectifier and then we have a low pass filter to follow the peak of the carrier. I have set the filter's cutoff frequency as 1000Hz. It is followed by a similar low pass filter to eliminate high frequency ripple.

The rectifier implemented here exploits the fact that an AM wave is symmetrical about the time axis. Also using a second LPF improves the performance of the AM receiver.

One can change the filter frequency characteristics to observe its effect on the output signal

4. **Amplify Output:** The demodulated wave can be small in amplitude. We can increase its signal level by editing the value in the block diagram.
5. **Save output file:** Enter the file location where you want the output demodulated wave to be saved. Save it with a .wav file extension.

The receiver front panel is as show below. It doesn't require any settings except for the file location for saving the output file as explained above.

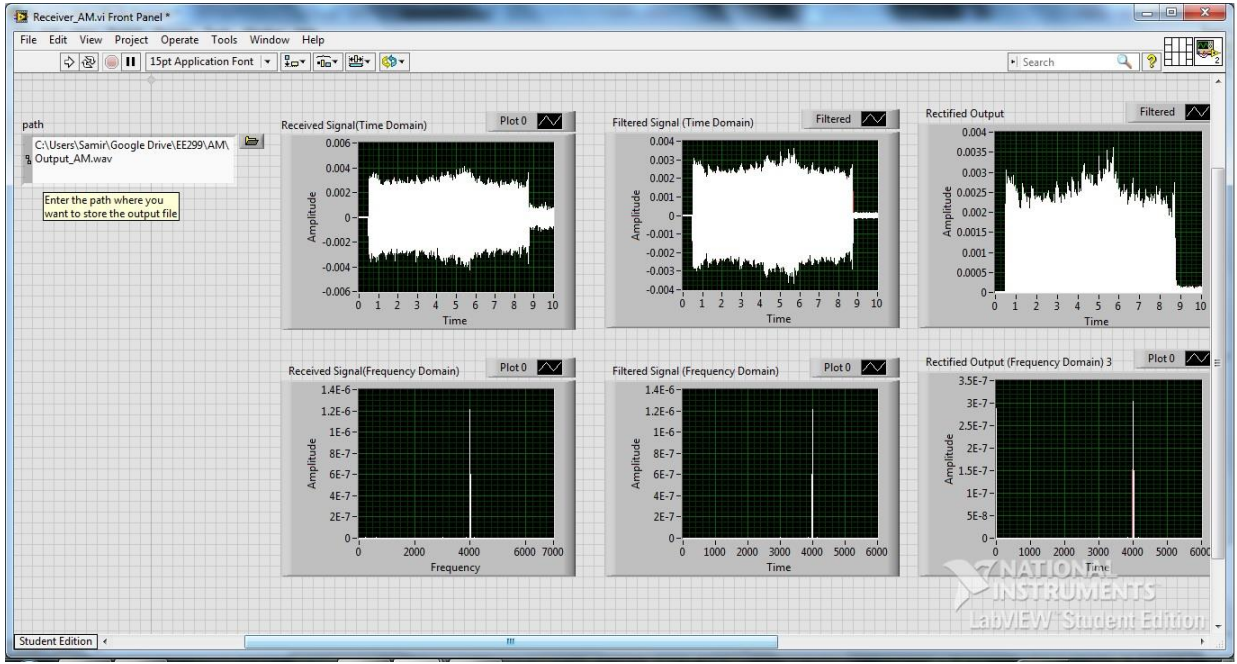

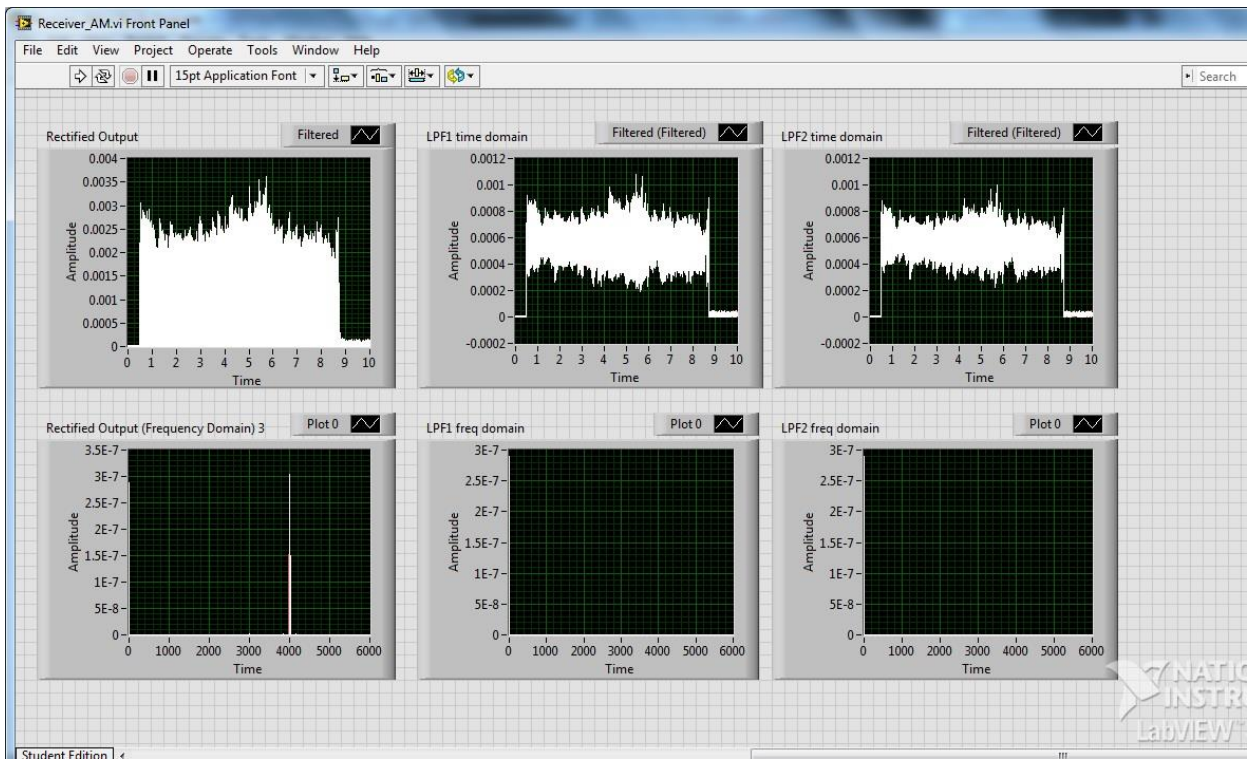

## Implementation:

1. Set the parameters as explained above in the transmitter and receiver blocks.
2. Keep the headphones near the microphone and set the volume to high.
3. Run Transmitter block and immediately Receiver block within a fraction of a second. Make sure you run the VIs only once and not continuously.
4. Listen to the audio output file saved on the disk.  Running the receiver block multiple times will overwrite the audio output file.
5. The scale of the graph can be changed by right clicking on it and selecting the options under the 'Scale' tab within 'Properties'. One can also simply right click and select 'Autoscale X-axis or Y-axis'. Another way to edit the scale by clicking on the specific value on the axis itself and changing it to the desired value.

# Frequency Modulation using LabVIEW

## User Guide

## Introduction

LabVIEW enables engineers to simulate various communication and control systems. LabVIEW helps to create Virtual Instruments (VIs) which are the files with which the user interacts to accomplish the required task.

The FM system is implemented using two separate VIs i.e. Transmitter_FM.vi and Receiver_AM.vi. Each VI has two parts: Front Panel and the Block Diagram. The Front Panel is usually the interface the user interacts with and observes results. The block diagram contains the blocks used to implement the functionality required for the operation of the VI. The individual blocks in the block diagram are called the sub VIs. The user may or may not need to make changes in the block diagram of the VI during the execution of the LabVIEW program.

## Hardware setup:

1. Make sure you have LabVIEW installed on your machine so that you can run the LabVIEW VIs on it.
2. Use a headphone as a transmitter for the FM wave.
3. The microphone of the machine will be used as the receiver. It is advisable to keep the headphone near the microphone for better results. It is also suggested to have a quiet environment during the whole experiment.

## FM Transmitter Working Principle

Consider a sinusoidal carrier wave defined by $c(t) = A_c \cos(2\pi f_c t)$ where $A_c$ is the carrier amplitude and $f_c$ is the carrier frequency.

Frequency modulation (FM) is that form of angle modulation in which the instantaneous frequency $f_i(t)$ is varied linearly with the message signal m(t), as shown by

$$f_i(t) = f_c + k_f m(t)$$

The constant term fc represents the frequency of the unmodulated carrier; the constant $k_f$ represents frequency-sensitivity factor of the modulator, expressed in hertz per volt. Integrating the above equation with respect to time and multiplying the result by 2π, we get

$$\theta_i(t) = 2\pi \int_0^t f_i(\tau) d\tau$$

$$= 2\pi f_c t + 2\pi k_f \int_0^t m(\tau) d\tau$$

Where the second term accounts for the increase or decrease in the instantaneous phase $\theta_i(t)$ due to the message signal m(t). The frequency-modulated wave is therefore

$$s(t) = A_c \cos[2\pi f_c t + 2\pi k_f \int_0^t m(\tau) d\tau]$$

FM LabVIEW User Guide

The following diagram shows how the frequency of the carrier wave changes according to the instantaneous amplitude of the message signal.

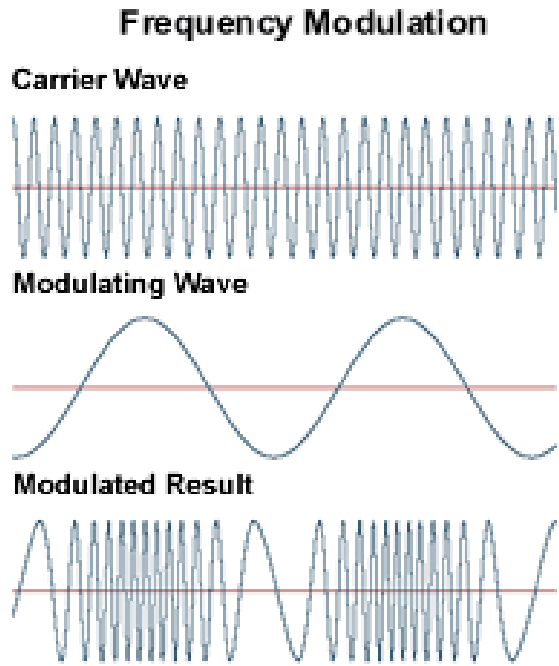

The frequency spectrum is seen as follows where the spectrum changes depending on the value of $k_f$ or M (in the figure). The center frequency corresponds to the carrier frequency.

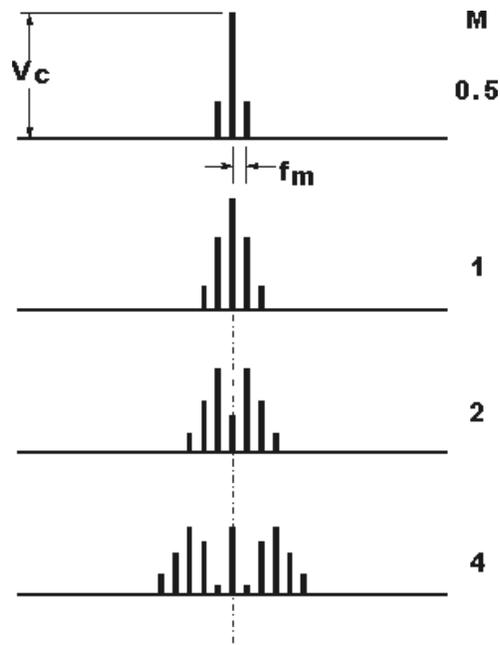

## Transmitter block

The transmitter block diagram looks as follows:

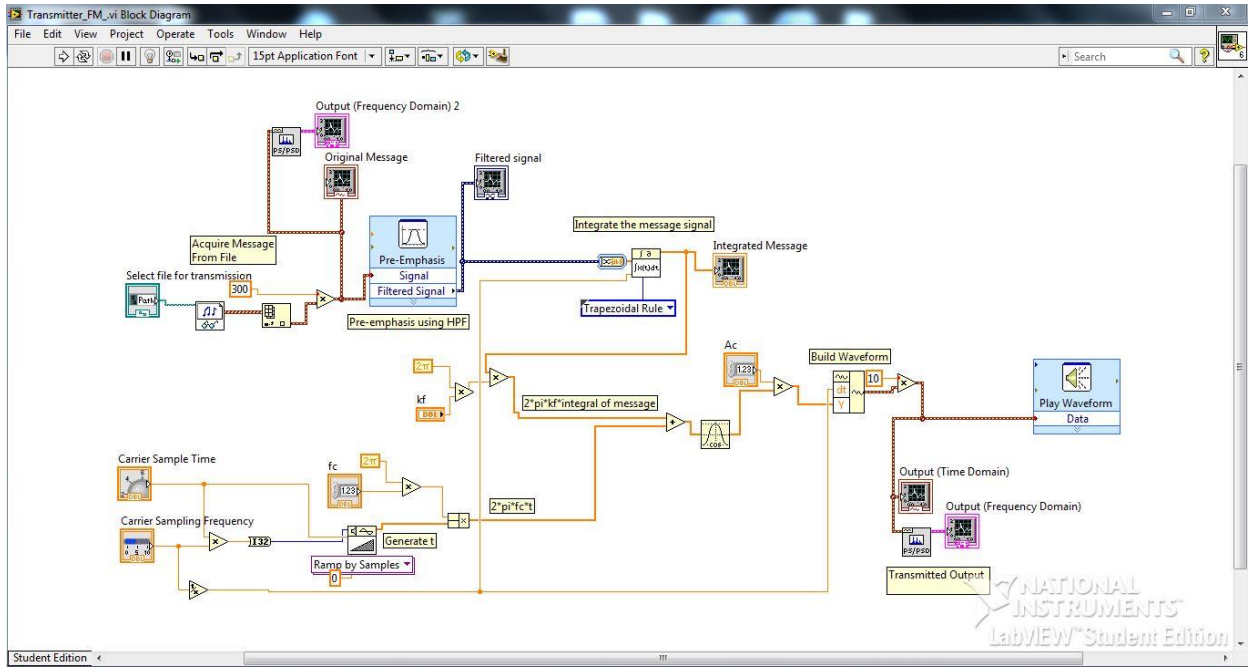

We all know that Frequency Modulation consists of changing the frequency of the high frequency carrier wave with a low frequency message signal. The above VI does the same by using a .wav file as a message.

The key functions in the block diagram are labeled in yellow. I will explain them one by one. Pay keen attention to the things to be edited in the front panel and the block diagram. Most of the blocks mentioned below are the only ones which can be used for that specific application in LabVIEW. Some suggestions for using different blocks are mentioned at particular places in the description below.

1.  **Generation of ($2\pi f_c t$):** This term is generated by using the sub VIs seen in the bottom left corner of the screenshot. The carrier sampling frequency is adjusted from the front panel to 44100 to be compatible with the sampling frequency of the usual audio files. The carrier sample time is set to 10 seconds to use the 10 second audio file for transmission. The 'Ramp Pattern' sub VI will generate the required 't' for the operation.
    **Tip:** Press 'Ctrl+H' to activate context help. Then whenever you take your cursor on any sub VI, it will briefly describe the input and the output pins of the sub VI in a small floating box.

2.  **Acquire Message from File:** In the space corresponding to this sub VI on the front panel, enter the path where the message file is stored. Make sure that the file is in the '.wav' format else LabVIEW will reject it.
    The 'Index Array' sub VI selects a sub-array of the entire n-dimensional array of the audio signal. In the block diagram, a constant is used for multiplication/amplification which can be edited. It can be replaced by a control, so that it can be adjusted from the front panel. To replace it, delete the constant input to it and right click on the sub VI input pin, and select 'create control' option.

**Tip:** Whenever we have to assign a path in LabVIEW, you don't have to type the entire path everything. You can right click on the path empty block and select Data Operations->Make Current Value Default.

3. **Pre-emphasis Block:** Pre-emphasis is an important concept which amplifies the high frequency components of the message signal and attenuates the low frequency components. An opposite operation called de-emphasis is carried on the receiver end. Pre-emphasis is done to improve the signal to noise ratio and for several other benefits.
We perform pre-emphasis by using a high pass filter with the cut-off frequency of 750Hz.

4. **Integration block:** The dynamic data of the message signal is converted to an array before integration. The rule for integration can be changed by clicking on the respective drop down arrow near the sub VI in the block diagram. We have employed the trapezoidal rule here. You can try different integration rules too. There are a few other integration blocks in LabVIEW but this one proved best for our application.

5. **PSD block:** It converts the time signal into a frequency spectrum data. It is displayed on the graph. Similar blocks can be observed in the entire block diagram. You can also add your own time signal graph blocks or PSD blocks anywhere in the block diagram to observe the waveform at that point of the system. Don't forget to position the corresponding graph on the front panel by dragging and adjusting its size.

6. **Build Waveform:** This sub VI builds the waveform from the output we got after applying the FM equation on the message signal. It requires the 'dt' value which is the reciprocal of the carrier sampling frequency.

7. **Power Amplify Output Wave:** Presently the output wave is power amplified by a factor of 10. You can change it if you want to by clicking in the small box in the block diagram.

**Tip:** The usual sampling frequency should be 44100 or 22050 because these values of sampling frequencies are compatible with most of the audio files and sound card drivers. Using these sampling values will prevent us from using resample waveform VI blocks in our design. The resample waveform block takes a signal and a 'dt' value as the inputs and it outputs the resampled waveform. The value 'dt' is the reciprocal of the sampling frequency desired.

**The front panel looks as follows**

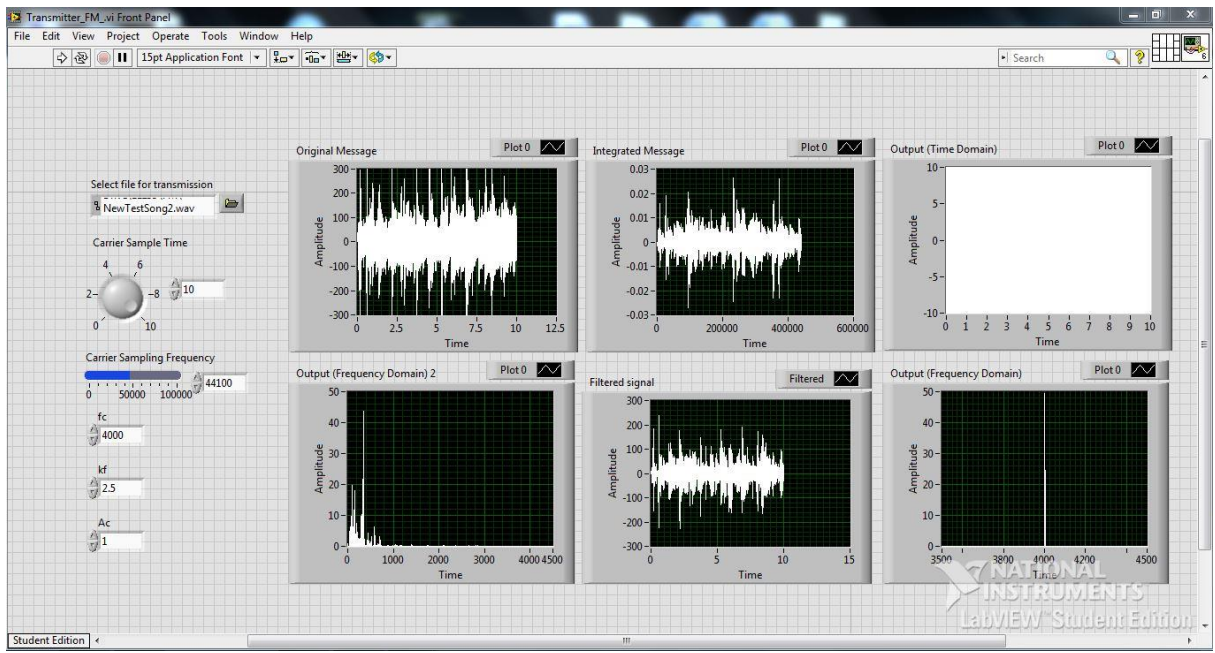

Most of the adjustments to be made in the front panel are explained above. My settings are 4kHz carrier with amplitude 1volt, $k_f$=2.5, carrier sample time =10 seconds for a 10 second message wave and 44100 as the carrier sampling frequency, so that the carrier is compatible with the sound card driver. You can try changing the value of kf to see the change in the frequency spectrum.

There is one more important setting to be set. It is in the 'Play Waveform' sub VI. When we use the 'Play Waveform' for the first time, you need to select the output device i.e. speakers or headphones. Test device to hear the audible tone. If you don't hear it, select another device. Rest of the settings fairly remains the same.

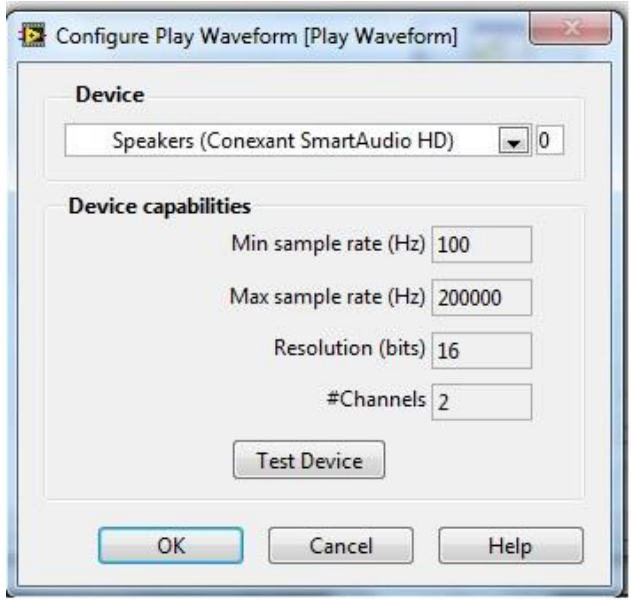

## FM Receiver Working Principle

Frequency demodulation is the process by means of which the original signal is recovered from an incoming FM wave. The frequency discriminator is used to demodulate a FM wave. A frequency discriminator consists of a slope detector followed by a envelope detector i.e. the reverse operation of the frequency modulator.

Consider the following FM wave

$$s(t) = A_c \cos[2\pi f_c t + 2\pi k_f \int_0^t m(\tau)d\tau]$$

Taking derivative, we get

$$\frac{ds(t)}{dt} = -2\pi A_c[f_c + k_f m(t)]\sin(2\pi f_c t + 2\pi k_f \int_0^t m(\tau)d\tau)$$

We observe that the derivative is a band-pass signal with amplitude modulation defined by the multiplying term$[f_c + k_f m(t)]$. Consequently, if $f_c$ is large enough such that the carrier is not overmodulated, then we can recover m(t) with an envelope detector. So the frequency discriminator consists of a differentiator followed by an envelope detector to recover the message signal.

## Receiver Block

The receiver block looks as shown below

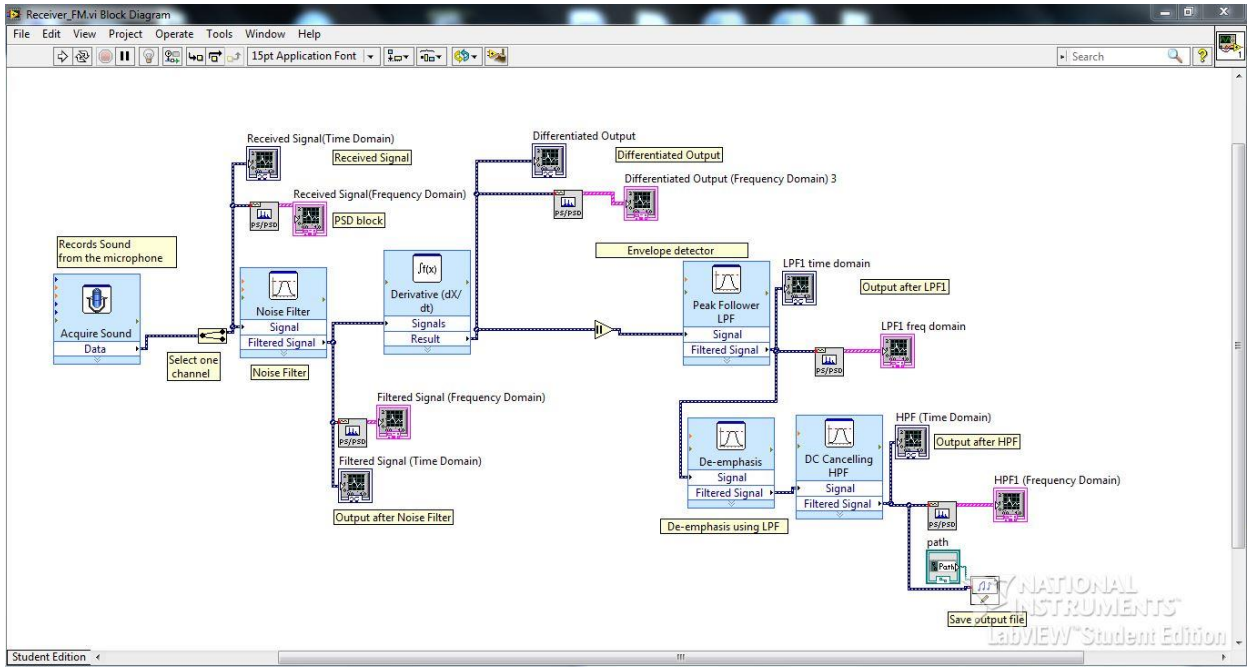

The important components labeled in yellow are explained below:

1. **Record Sound from the Microphone:** This sub VI is used to record the sound from the microphone.

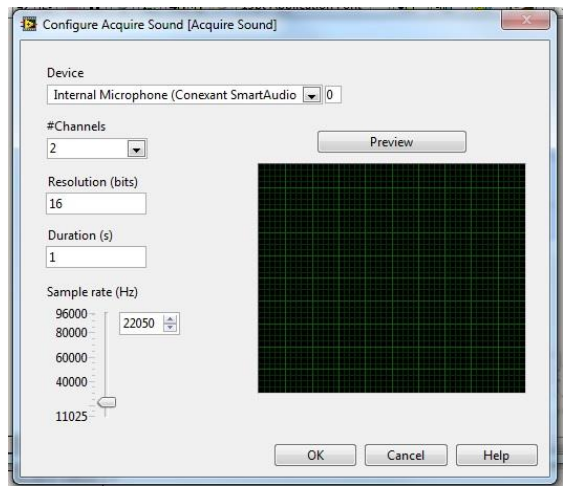

Select the input device i.e. microphone which you are using and also specify the number of seconds you want to record the sound for. For an input message signal of 10 seconds, you should keep the duration in the 'Acquire Sound' sub VI as 10 seconds too. Next we select only one channel of the audio using the 'Split Signals' sub VI.

2. **Noise Filter:** Filters are used extensively in all communication system implementations in LabVIEW. This is how the filter sub VI looks like.

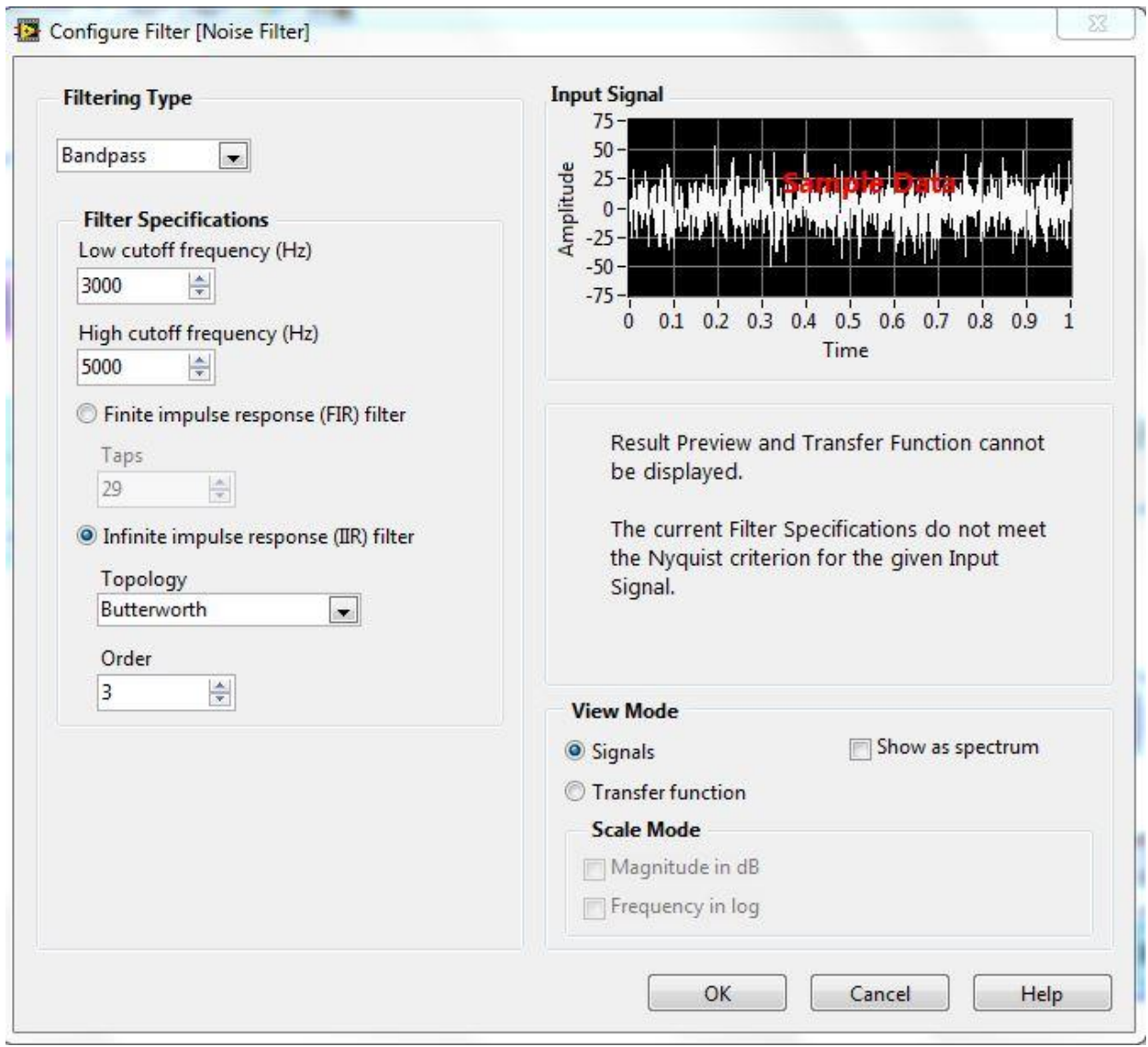

You can select the filter type and the cutoff frequency as well as the order of the filter. I used a bandpass filter with a bandwidth of 2kHz around the carrier frequency of 4kHz to eliminate noise and because the message sound file had a spectrum from 0Hz to 1000Hz.

3.  **Derivative block:** This sub VI takes the derivative of the signal. This sub VI worked the best for our application; else there are a few other derivative blocks in LabVIEW.
4.  **Envelope Detector block and other Filters:** We know how the AM peak detector circuit or an envelope detector looks like. The diode acts as a rectifier and then we have a low pass filter to follow the peak of the carrier. I have set the filter's cutoff frequency as 500Hz. It is followed by a low pass filter with a cutoff frequency of 750 Hz to perform de-emphasis i.e. attenuation of the high frequency components which were amplified in the transmitter's pre-emphasis operation. The high pass filter of cutoff frequency 1000Hz enhances the performance by eliminating the DC component in the received signal. One can easily observe the high DC content by observing the time domain signal before the high pass filter.
    One can change the filter frequency characteristics to observe its effect on the output signal. The filter parameters selected are optimum for frequency demodulation in our case.
5.  **Amplify Output(Optional):** The demodulated wave can be small in amplitude. We can increase its signal level by multiplying the signal by a constant in the block diagram.

6. **Save output file:** Enter the file location where you want the output demodulated wave to be saved. Save it with a .wav file extension.

The receiver front panel is as show below. It doesn't require any settings except for the file location for saving the output file as explained above.

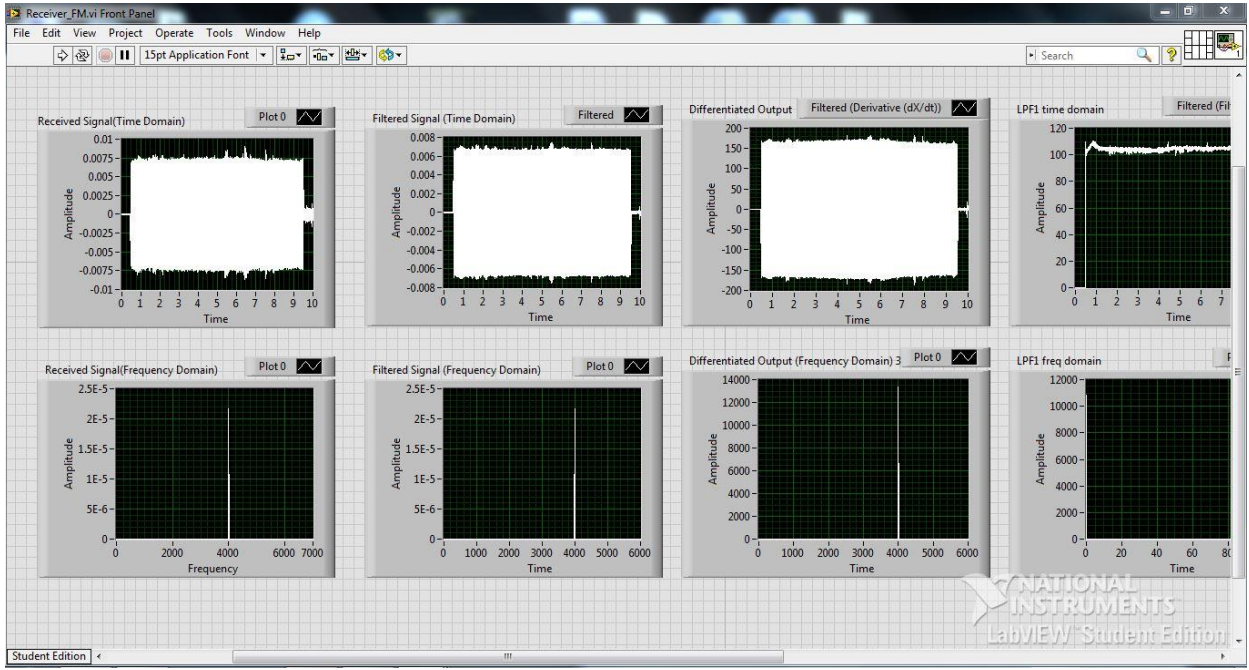

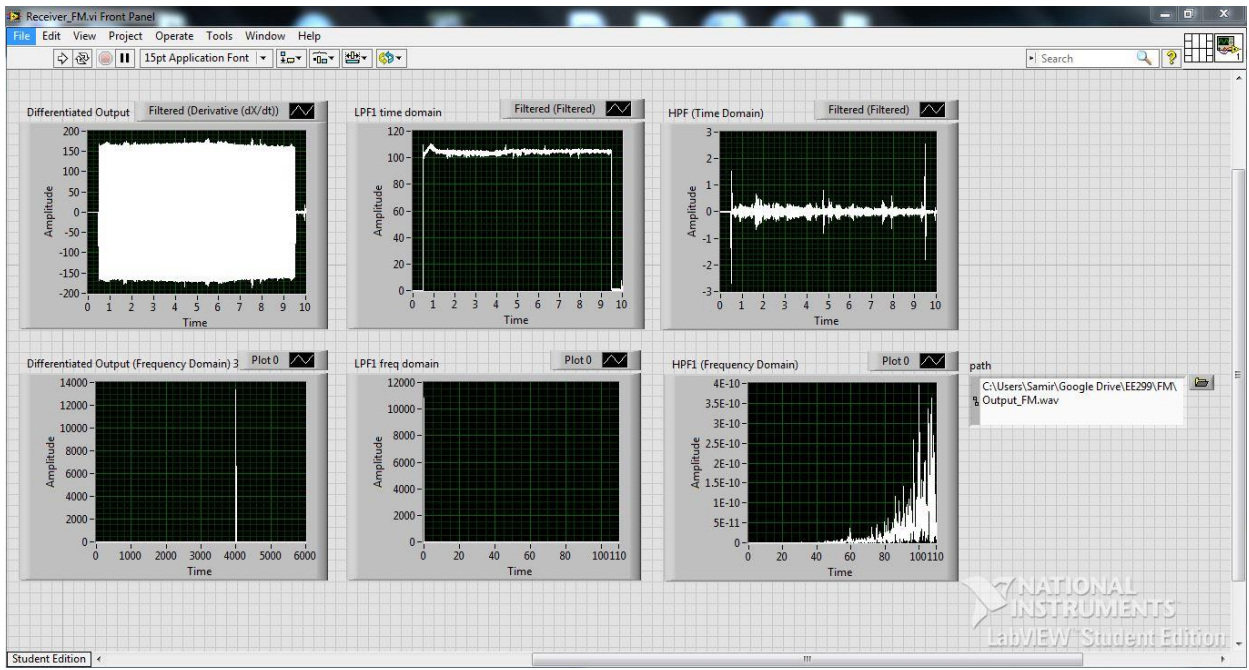

## Implementation:

1. Set the parameters as explained above in the transmitter and receiver blocks.
2. Keep the headphones near the microphone and set the volume to high.

3. Run the Transmitter block and immediately the Receiver block within a fraction of a second. Make sure you run the VIs only once and not continuously.
4. Listen to the audio output file saved on the disk. Running the receiver block multiple times will overwrite the audio output file.
5. The scale of the graph in the front panel can be changed by right clicking on it and selecting the options under the 'Scale' tab within 'Properties'. One can also simply right click and select 'Autoscale X-axis or Y-axis'. Another way to edit the scale by clicking on the specific value on the axis itself and changing it to the desired value.

# Binary Frequency Shift Keying using LabVIEW

## User Guide

## Introduction

LabVIEW enables engineers to simulate various communication and control systems. LabVIEW helps to create Virtual Instruments (VIs) which are the files with which the user interacts to accomplish the required task.

The BFSK system is implemented using two separate VIs i.e. BFSK_Transmitter.vi and BFSK_Receiver.vi. Each VI has two parts: Front Panel and the Block Diagram. The Front Panel is usually the interface the user interacts with and observes results. The block diagram contains the blocks used to implement the functionality required for the operation of the VI. The individual blocks in the block diagram are called the sub VIs. The user may or may not need to make changes in the block diagram of the VI during the execution of the LabVIEW program.

## Hardware setup:

1. Make sure you have LabVIEW installed on your machine so that you can run the LabVIEW VIs on it.
2. No other hardware like a microphone or a speaker is required because we are using a file based system wherein the transmitter's output is written onto a file and contents are read from the same file by the receiver.

## BFSK Transmitter Working Principle

Binary frequency shift keying (BFSK) is a simple form of digital modulation where the amplitude and the phase of the carrier remain constant. The output is dependent on the symbols 0 and 1 to be transmitted. When the symbol to be transmitted is 0, the carrier $c1(t)=A_c \cos(2\pi f_1 t)$ is transmitted during that symbol interval. The carrier $c2(t)=A_c \cos(2\pi f_2 t)$ is transmitted when the transmitted symbol is 1.

The following diagram shows how a BFSK wave is modulated by a digital signal

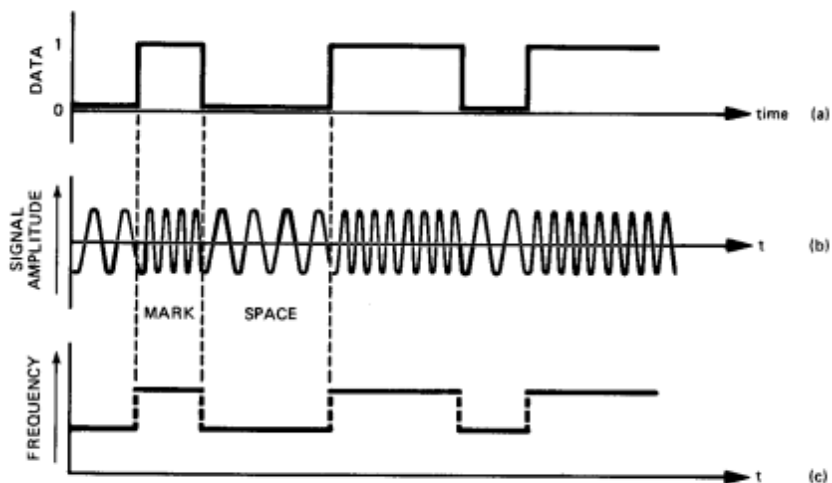

Figure 1. FSK modulation. Binary data (a) frequency modulates the carrier to produce the FSK signal (b) which has the frequency characteristic (c).

## Transmitter block

The transmitter block diagram looks as follows:

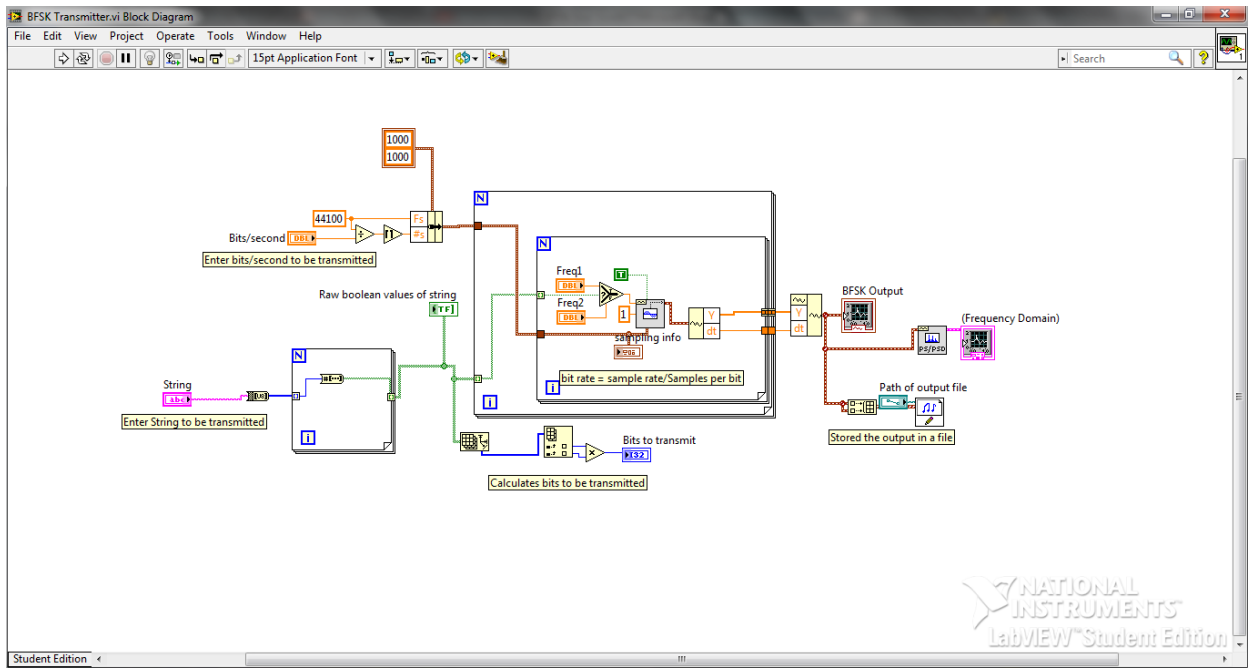

The key functions in the block diagram are labeled in yellow. I will explain them one by one. Pay keen attention to the things to be edited in the front panel and the block diagram. Most of the blocks mentioned below are the only ones which can be used for that specific application in LabVIEW. Some suggestions for using different blocks are mentioned at particular places in the description below.

1. **String to be transmitted:** Enter the string to be transmitted in the front panel. Here we use a single letter. The blocks in the lower left of the block diagram convert each letter into a 8 bit byte array. 'String to Byte' sub VI is used to convert the string to a byte array. Then using 'Number to Boolean' sub VI in a for loop, each bit is stored in a Boolean array. The small box indicating the array entering and exiting the for loop is used to auto-index the array by default. We can change this setting by right-clicking on the small box. We don't need to disable/change the auto-indexing option in our application. Subsequently the bits to be transmitted are displayed on the front panel depending on the number of elements in the array generated. The 'Array Size' sub VI helps us to obtain the length of the array and hence the number of bits to be transmitted.

2. **Symbol Interval:** The top left of the block diagram shows the variable for the bits/second. The bits/second is set in the front panel. Suppose a letter 'h' is to be transmitted and it is represented by 8 bits and if bits/second is equal to 1, then the output waveform is of 8 seconds.

3. **For loop for BFSK operation:** The string array converted to a binary array is usually a 2-D array. The first for loop is used to extract each row of the array; hence it selects a 1-D array. The second for loop is used to extract each bit from the 1-D array obtained from the first for loop. It then selects frequency of the output waveform (using the select sub VI) to be either of the two frequencies entered by the user in the front panel. In our case, the frequencies selected are 4kHz and 6kHz, which can be adjusted too. The sampling frequency is set to 44100 in the block diagram by the user (see top left corner of the block diagram), which can be adjusted as per requirements. This sampling frequency value enters the for loops without auto-indexing since it is a single value and not an array.

Finally the 'Get Waveform Components' sub VI and the 'Build Waveform' VI generates the required output wave.

4. **Graphs:** The time domain graph of the waveform is shown using the waveform graph sub VI. The PSD block converts the time signal into a frequency spectrum data. It is displayed on the graph. You can also add your own time signal graph blocks or PSD blocks wherever in the block diagram to observe the waveform at that point of the system. Don't forget to position the corresponding graph on the front panel by dragging and adjusting its size.

5. **Save output on file:** The array of the output waveform is built using the 'Build Array' sub VI. This array is written into a sound file on the disk at the path specified in the 'Sound file write simple' sub VI. Make sure to name the file with a .wav extension. You can change the path by making the required changes in the front panel. You can also make the value in the path box default, so that you don't have to change it frequently. To do that, right click on the box and select Data Operations>>Make Current Value Default. Also multiple writes to the same file rewrites the original file.

**The front panel looks as follows**

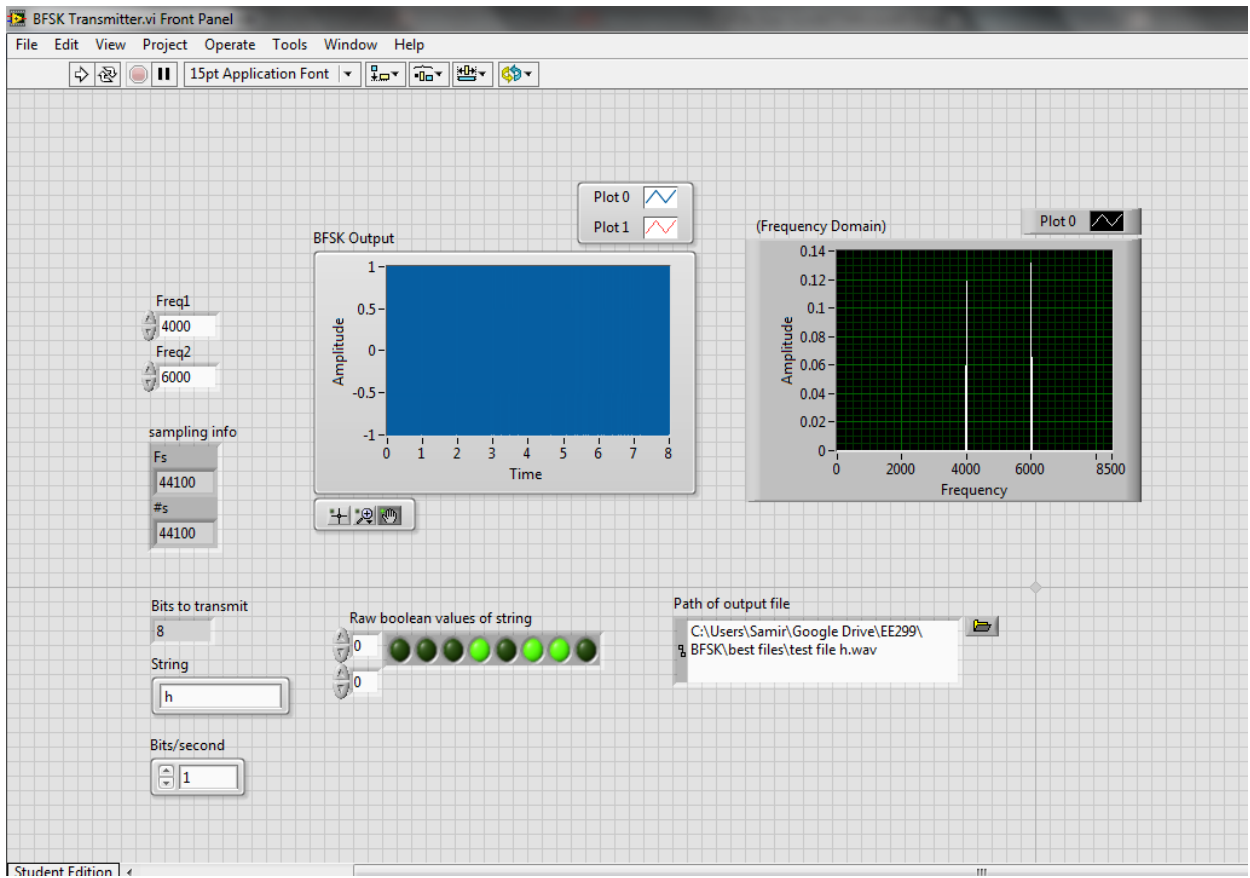

## BFSK Receiver Working Principle

The diagram below shows the working of a BFSK non-coherent receiver.

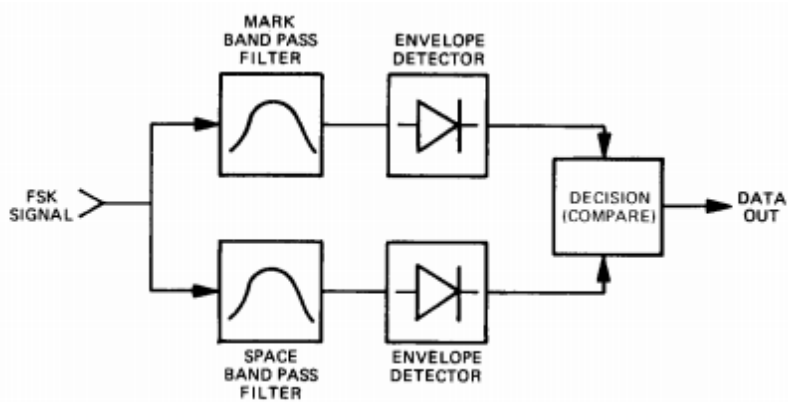

The band pass filters are having their center frequencies set to the two frequencies used for communication. Hence the output of each band pass filter will have the signal content corresponding to that particular frequency along with some noise in the pass band. Then the signal is passed through an envelope detector which is a diode followed by a low pass filter. Finally the signal levels from both the branches in the symbol period are compared and the branch from which the stronger signal level is received, that symbol (0 or 1) is selected for that symbol period. The above procedure is followed till all the symbols are decoded in the BFSK waveform.

## Receiver Block

The receiver block looks as shown below

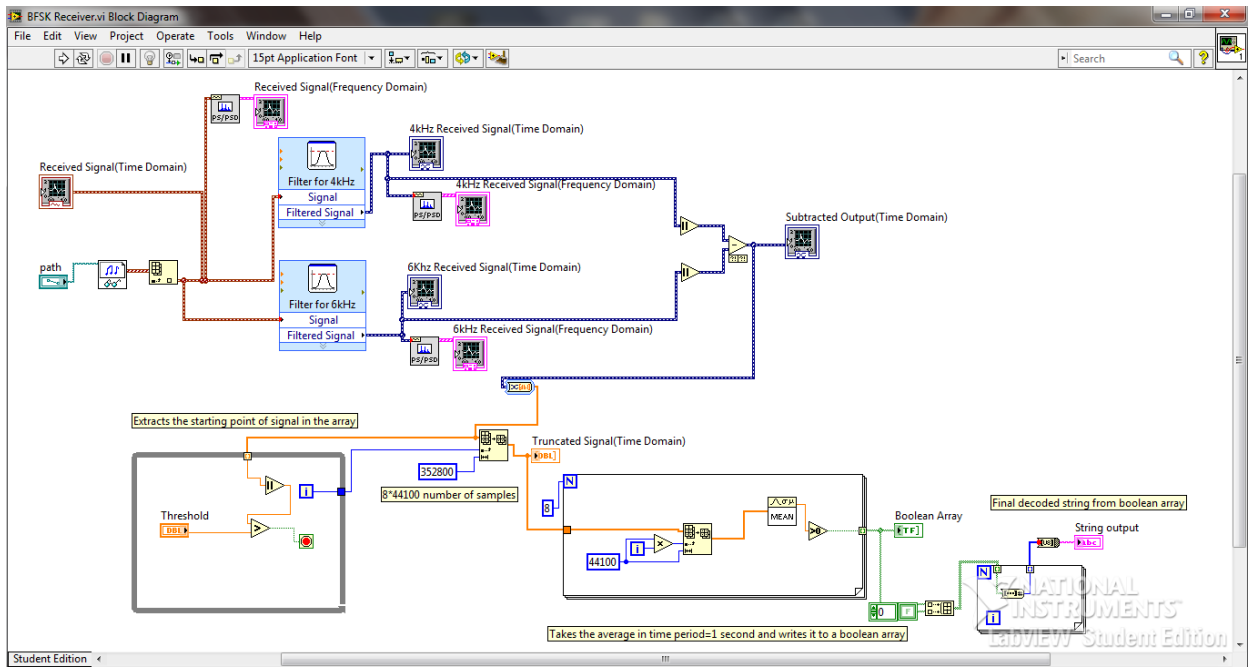

The important components labeled in yellow are explained below:

1.  **Read sound from the file:** The 'Sound file read simple' sub VI is used to read the file from the disk for demodulation. Select the file from the front panel by clicking and selecting the file from the path section. The 'Index Array' sub VI is used to produce the sound input array for demodulation.

2. **Band Pass Filters:** Filters are used extensively in all communication system implementations in LabVIEW. This is how the filter sub VI looks like.

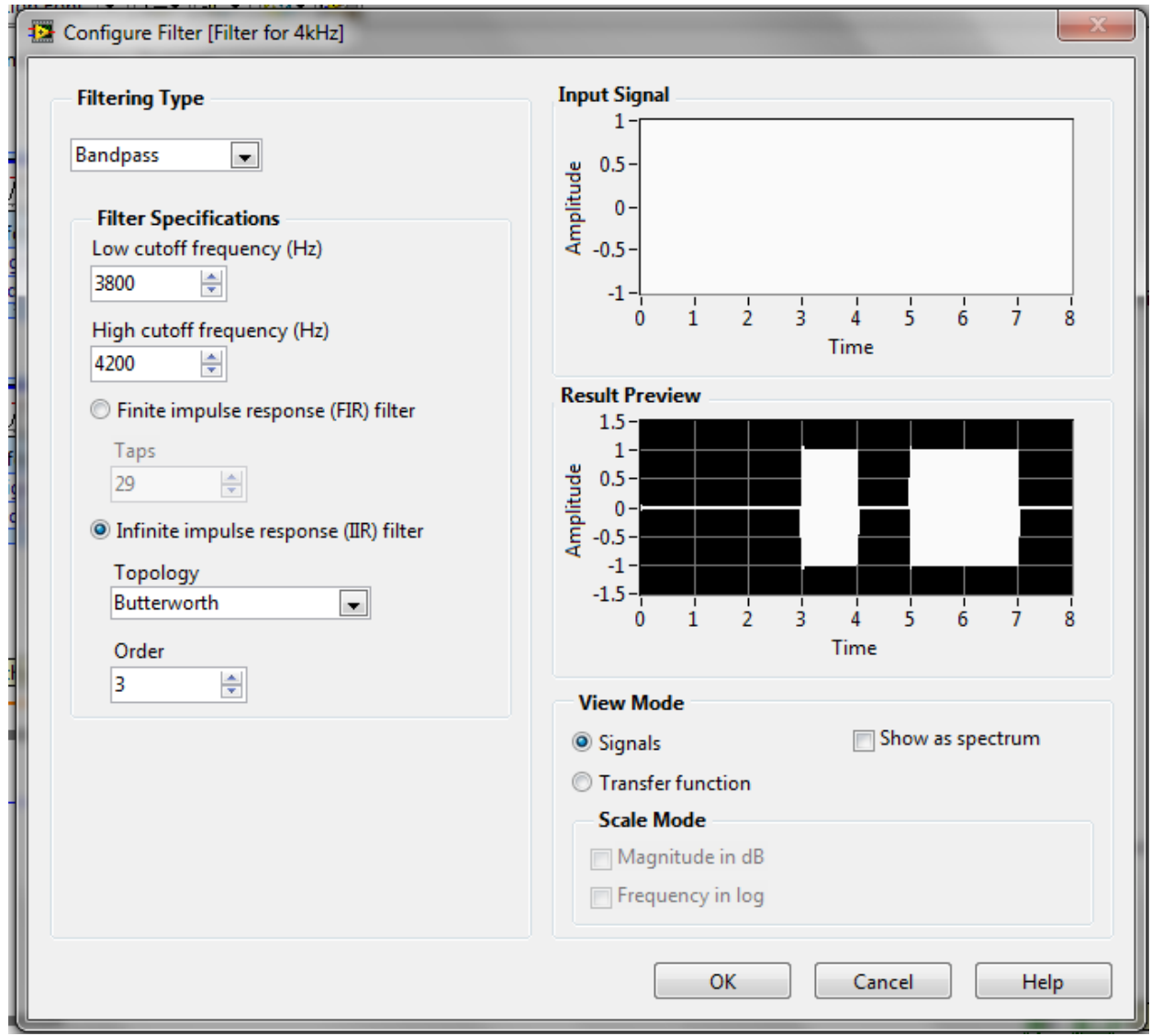

You can select the filter type and the cutoff frequency as well as the order of the filter. I used a bandpass filter with a bandwidth of 400 Hz around the carrier frequency of 4kHz to eliminate noise.

3. **Rectifier block and Filters:** We know how an envelope detector circuit looks like. The diode acts as a rectifier and then we have a low pass filter to follow the peak of the carrier. We take the modulus of the signal to produce an effect like a diode being used. We don't need a low pass filter in our application.

4. **Extract starting point:** This step is necessary if the wave is transmitted over the air medium. We should know the exact instant when the actual signal starts, rest is noise. Hence we set a threshold above which the signal is the legitimate signal. The threshold can be changed from the front panel.

First we convert the output from the filter (dynamic data) to an array. Then we run a while loop with the stop condition which stops the loop when the signal's absolute value is greater than the threshold. This helps us extract the exact starting point of the signal

5. **Get truncated signal:** Once the starting point is extracted, we need only those number of points, after that starting point, which contain the signal. Here we know there are 44100 points per second and for a 8 second BFSK waveform, the total number of points will be 44100*8. Thus we get a truncated waveform from the 'Array Subset' waveform.

6. **Decode the signal:** We have a for loop which runs 8 times because we have to decode 8 symbol periods. The sub VIs in the for loop take the mean of every 44100 samples of the array and output 0 or 1 by comparing the calculated mean with zero. Lastly the string is decoded and displayed on the front panel. A few data type conversions are necessary before the final conversion. All those conversions are self-explanatory when you activate context help (press Ctrl+H) and then take the mouse cursor over those sub VIs.

The receiver front panel is as shown below. It doesn't require any settings except for the file location for selecting the BFSK waveform file and the threshold value selection.

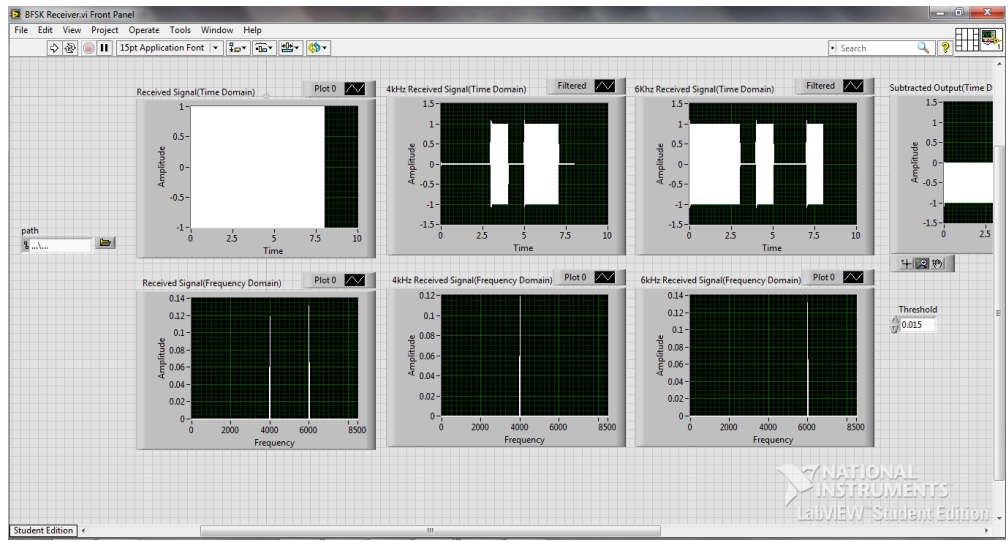

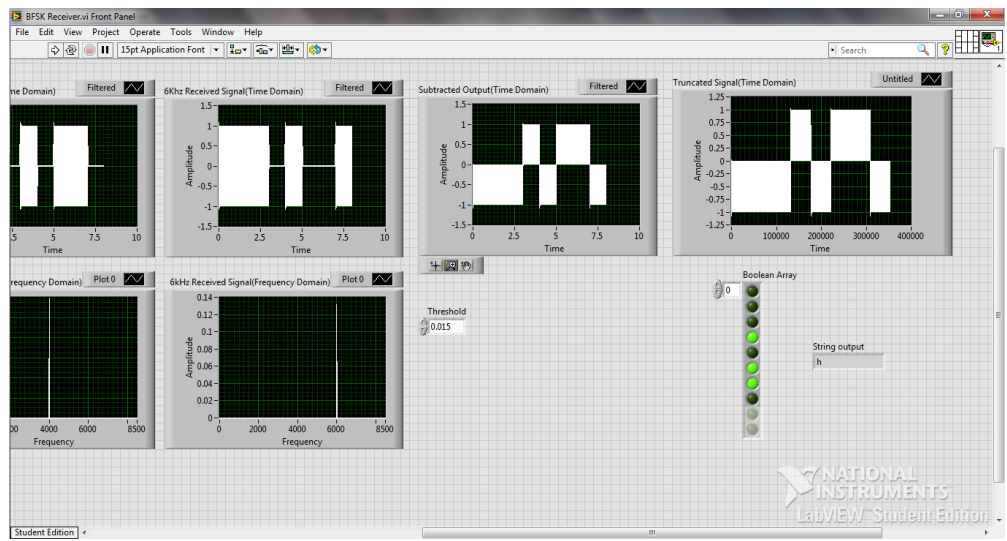

## Implementation:

1. Set the parameters as explained above in the transmitter and receiver blocks.
2. Run Transmitter block and it will save the output file.
3. Run the Receiver block by using the saved file.
4. The scale of the graph can be changed by right clicking on it and selecting the options under the 'Scale' tab within 'Properties'. One can also simply right click and select 'Autoscale X-axis or Y-axis'. Another way to edit the scale by clicking on the specific value on the axis itself and changing it to the desired value.

**Notes for future design:** Later versions of this design can include transmission using the air medium or by using the NI myDAQ. While using the air for transmission, the microphone works as the receiver. Some laptops have advanced noise filtering microphone which sometimes hinder the process of capturing single-tone or similar transmitted waves. Hence the NI myDAQ can prove useful in such situations. Either the microphone should be used on the NI myDAQ or the data can just be written to it and read back, just like we did in this LabVIEW design. While transmitting over the air, selecting the threshold is a tricky part. We need to observe the waveforms to determine the right value of the threshold.

# Quadrature Amplitude Modulation using LabVIEW

User Guide

## Introduction

LabVIEW enables engineers to simulate various communication and control systems. LabVIEW helps to create Virtual Instruments (VIs) which are the files with which the user interacts to accomplish the required task.

The BFSK system is implemented using two separate VIs i.e. QAM Transmitter.vi and QAM Receiver.vi. Each VI has two parts: Front Panel and the Block Diagram. The Front Panel is usually the interface the user interacts with and observes results. The block diagram contains the blocks used to implement the functionality required for the operation of the VI. The individual blocks in the block diagram are called the sub VIs. The user may or may not need to make changes in the block diagram of the VI during the execution of the LabVIEW program.

## Hardware setup:

1. Make sure you have LabVIEW installed on your machine so that you can run the LabVIEW VIs on it.
2. No other hardware like a microphone or a speaker is required because we are using a file based system wherein the transmitter's output is written onto a file and contents are read from the same file by the receiver.

## QAM Transmitter Working Principle

Quadrature amplitude modulation (QAM) or quadrature-carrier multiplexing is a form of digital modulation where two messages are transmitted by modulating two waves of the same frequency but both carrier waves being 90 degrees apart in phase. The equation of a QAM wave is given below where $m_1(t)$ and $m_2(t)$ are the two messages.

$$s(t)=m_1(t)*A_c \cos(2\pi f_1 t) + m_2(t)*A_c \sin(2\pi f_1 t)$$

The following diagram shows how a QAM wave is modulated by digital signals $m_1(t)$ and $m_2(t)$

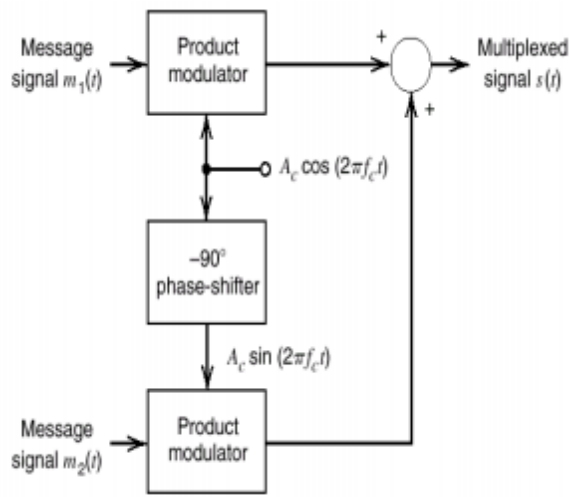

# Transmitter block

The transmitter block diagram looks as follows:

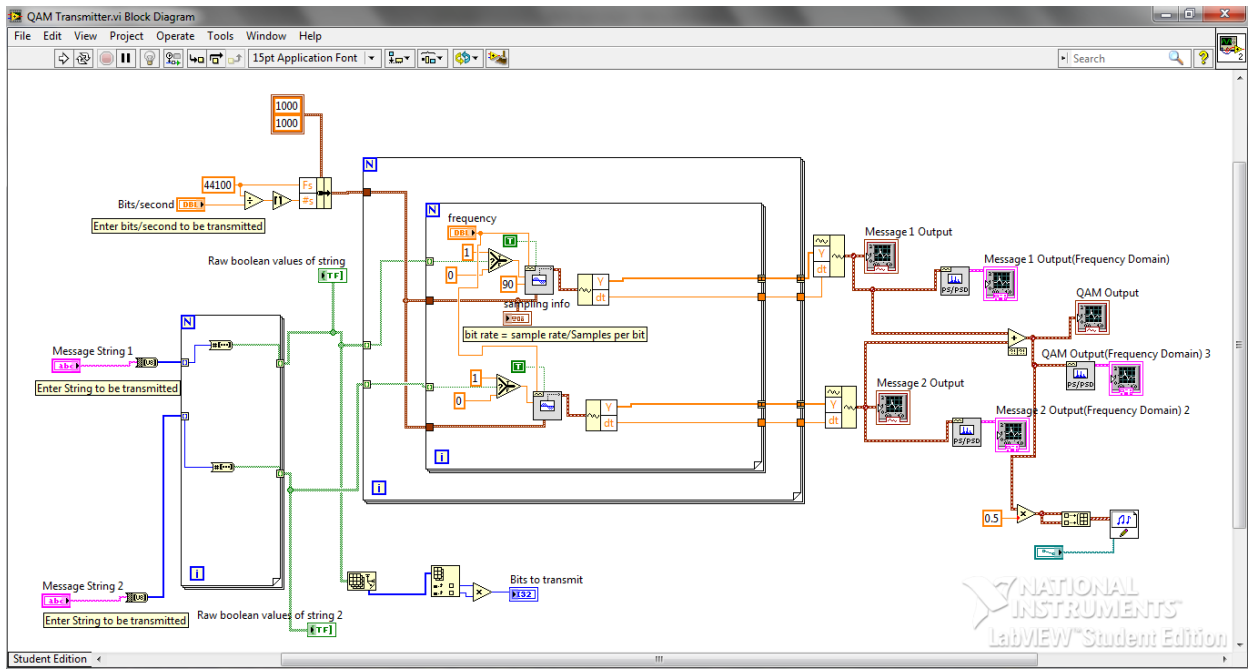

The key functions in the block diagram are labeled in yellow. I will explain them one by one. Pay keen attention to the things to be edited in the front panel and the block diagram. Most of the blocks mentioned below are the only ones which can be used for that specific application in LabVIEW. Some suggestions for using different blocks are mentioned at particular places in the description below.

The set of operations on each message is the same. Only the carrier of one is 90 degrees apart from the other one i.e. the 'Sine Waveform' sub VI has a 90 degrees phase offset for generating a cosine wave while the other one generate a sine wave with no phase offset.

1.  **String to be transmitted:** Enter the string to be transmitted in the front panel. Here we use a single letter. The blocks in the lower left of the block diagram convert each letter into a 8 bit byte array. 'String to Byte' sub VI is used to convert the string to a byte array. Then using 'Number to Boolean' sub VI in a for loop, each bit is stored in a Boolean array. The small box indicating the array entering and exiting the for loop is used to auto-index the array by default. We can change this setting by right-clicking on the small box. We don't need to disable/change the auto-indexing option in our application. Subsequently the bits to be transmitted are displayed on the front panel depending on the number of elements in the array generated. The 'Array Size' sub VI helps us to obtain the length of the array and hence the number of bits to be transmitted.

2.  **Symbol Interval:** The top left of the block diagram shows the variable for the bits/second. The bits/second is set in the front panel. Suppose a letter 'h' is to be transmitted and it is represented by 8 bits and if bits/second is equal to 1, then the output waveform is of 8 seconds.

3.  **For loop for BFSK operation:** The string array converted to a binary array is usually a 2-D array. The first for loop is used to extract each row of the array; hence it selects a 1-D array. The second for loop is used to extract each bit from the 1-D array obtained from the first for loop. It then selects the amplitude to be 0 or 1 (using the select sub VI) depending on the bits present in the bits array. In our case, the frequency selected is 4kHz which can be adjusted too from the front panel. The sampling frequency is set to 44100 in the block diagram by the user (see top left corner of the block diagram), which can be adjusted as per requirements. This sampling frequency value enters the for loops without auto-indexing since it is a single value and not an array.

Finally the 'Get Waveform Components' sub VI and the 'Build Waveform' VI generates the required output wave.

4. **Graphs:** The time domain graph of the waveform is shown using the waveform graph sub VI. The PSD block converts the time signal into a frequency spectrum data. It is displayed on the graph. You can also add your own time signal graph blocks or PSD blocks wherever in the block diagram to observe the waveform at that point of the system. Don't forget to position the corresponding graph on the front panel by dragging and adjusting its size.

5. **Save output on file:** The array of the output waveform is built using the 'Build Array' sub VI. This array is written into a sound file on the disk at the path specified in the 'Sound file write simple' sub VI. Make sure to name the file with a .wav extension. You can change the path by making the required the changes in the front panel. You can also make the value in the path box default, so that you don't have to change it frequently. To do that, right click on the box and select Data Operations>>Make Current Value Default. Also multiple writes to the same file rewrites the original file. The wave is multiplied by 0.5 in the end to avoid it going above +1 or below -1 in amplitude. This is crucial if air is used as the medium transmission but not in this design.

**The front panel looks as follows**

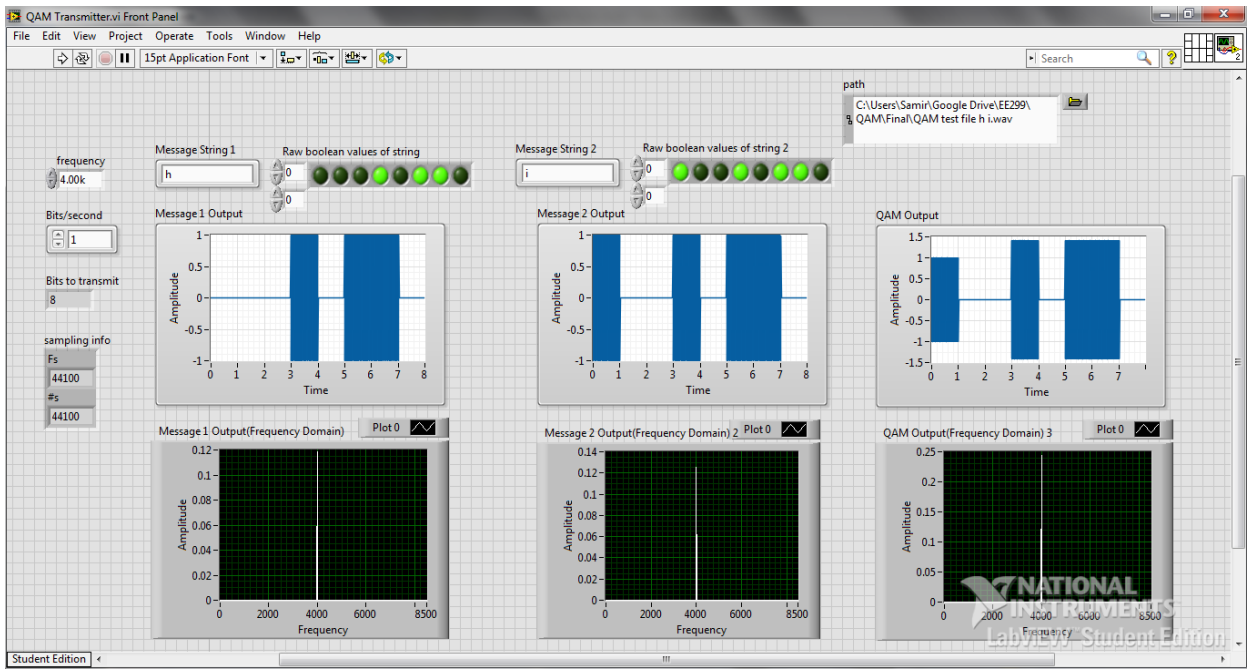

# QAM Receiver Working Principle

The diagram below shows the working of a QAM receiver.

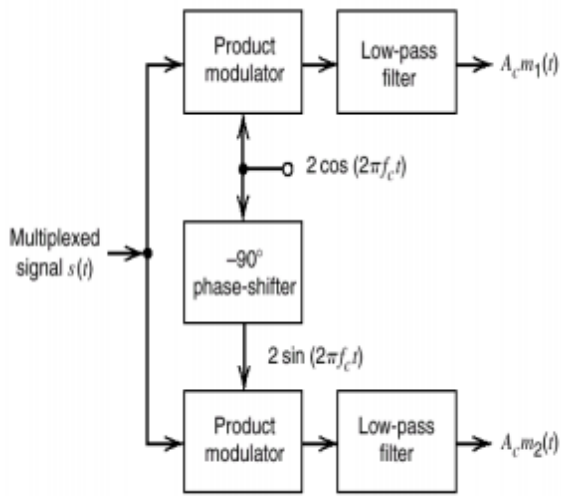

The diagram is self-explanatory. The locally generated sine and cosine wave are multiplied with the incoming QAM wave. The low pass filters are used to remove the higher frequency harmonics of the main frequency. The low frequency message signals are obtained at the output.

# Receiver Block

The receiver block looks as shown below

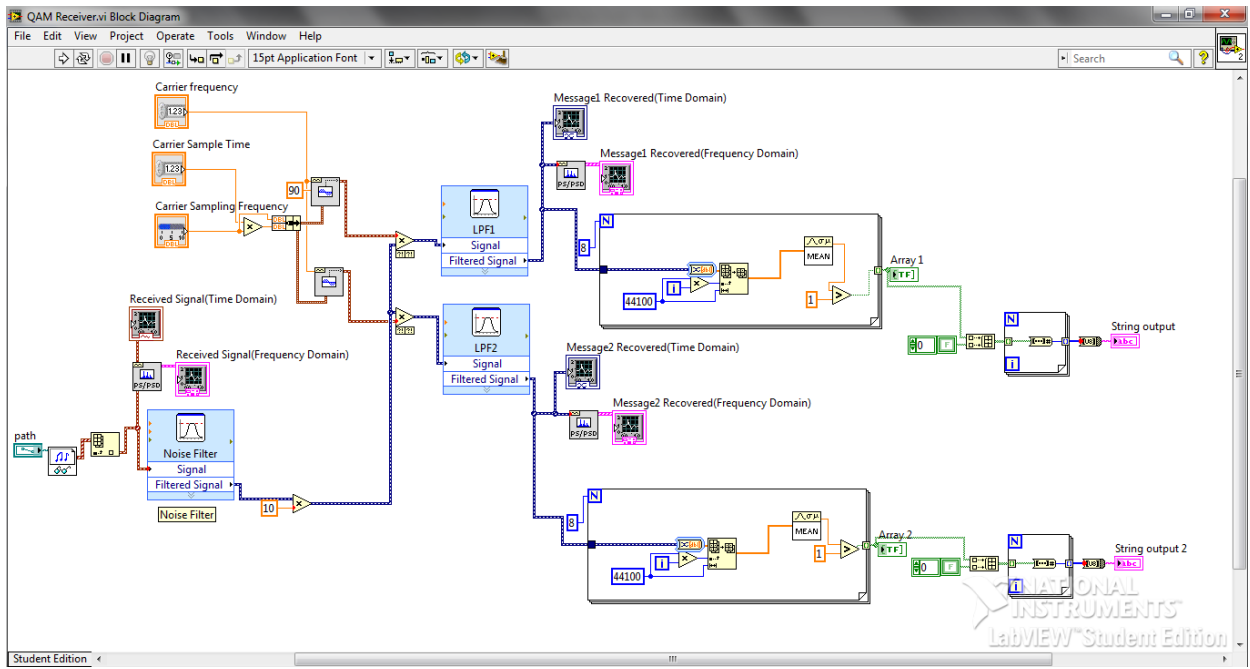

The important components labeled in yellow are explained below:

1. **Read sound from the file:** The 'Sound file read simple' sub VI is used to read the file from the disk for demodulation. Select the file from the front panel by clicking and selecting the file from the path section. The 'Index Array' sub VI is used to produce the sound input array for demodulation.

2. **Band Pass Filters:** Filters are used extensively in all communication system implementations in LabVIEW. This is how the filter sub VI looks like.

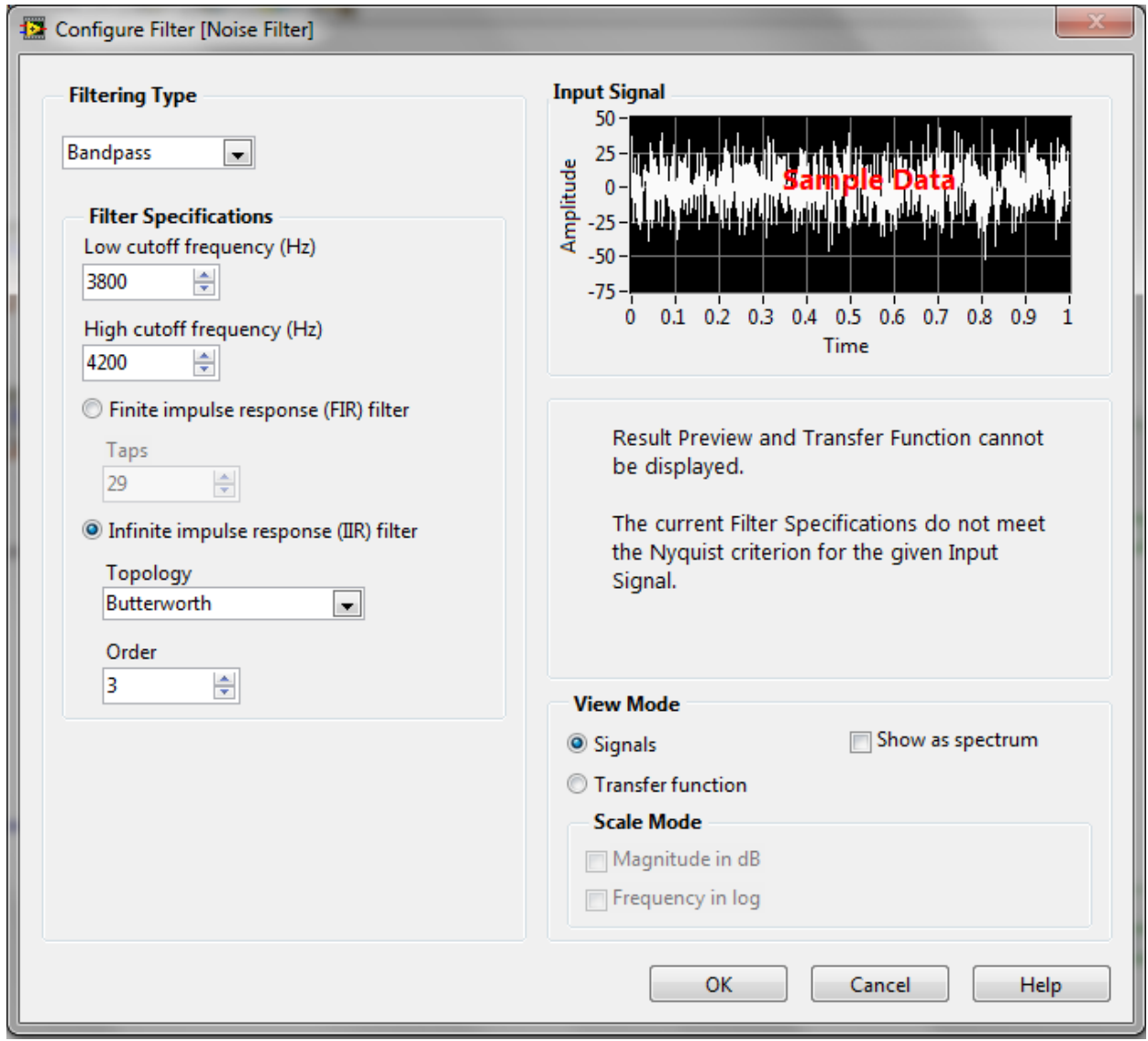

You can select the filter type and the cutoff frequency as well as the order of the filter. I used a bandpass filter with a bandwidth of 400 Hz around the carrier frequency of 4kHz to eliminate noise. The low pass filters have a cut off frequency of 5kHz.

3. **Rectifier block and Filters:** We know how an envelope detector circuit looks like. The diode acts as a rectifier and then we have a low pass filter to follow the peak of the carrier. We take the modulus of the signal to produce an effect like a diode being used. We don't need a low pass filter in our application.

**Note:** Extraction of starting point: This step is necessary if the wave is transmitted over the air medium. We should know the exact instant when the actual signal starts, rest is noise. Hence we set a threshold above which the signal is the legitimate signal. The threshold can be changed from the front panel.First we convert the output from the filter (dynamic data) to an array. Then we run a while loop with the stop condition which stops the loop when the signal's absolute

value is greater than the threshold. This helps us extract the exact starting point of the signal. We don't need this feature in this design.

4. **Decode the signal:** We have a for loop which runs 8 times because we have to decode 8 symbol periods. The sub VIs in the for loop take the mean of every 44100 samples of the array and output 0 or 1 by comparing the calculated mean with zero. Lastly the string is decoded and displayed on the front panel. A few data type conversions are necessary before the final conversion. All those conversions are self-explanatory when you activate context help (press Ctrl+H) and then take the mouse cursor over those sub VIs.

The receiver front panel is as shown below. It doesn't require any settings except for the file location for selecting the BFSK waveform file and the threshold value selection.

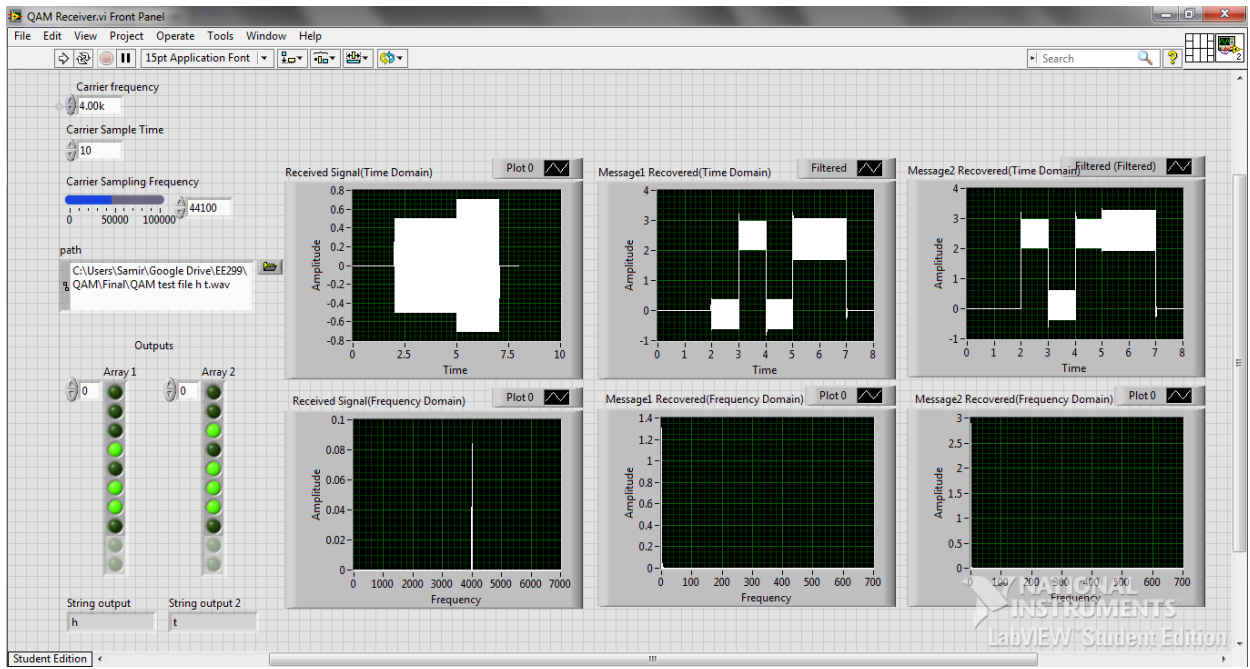

## Implementation:

1. Set the parameters as explained above in the transmitter and receiver blocks.
2. Run Transmitter block and it will save the output file.
3. Run the Receiver block by using the saved file.
4. The scale of the graph can be changed by right clicking on it and selecting the options under the 'Scale' tab within 'Properties'. One can also simply right click and select 'Autoscale X-axis or Y-axis'. Another way to edit the scale by clicking on the specific value on the axis itself and changing it to the desired value.

**Notes for future design:** Later versions of this design can include transmission using the air medium or by using the NI myDAQ. While using the air for transmission, the microphone works as the receiver. Some laptops have advanced noise filtering microphone which sometimes hinder the process of capturing single-tone or similar transmitted waves. Hence the NI myDAQ can prove useful in such situations. Either the microphone should be used on the NI myDAQ or the data can just be written to it and read back, just like we did in this LabVIEW design. While transmitting over the air, selecting the threshold is a tricky part. We need to observe the waveforms to determine the right value of the threshold.